\documentclass[fleqn,usenatbib,useAMS]{mnras}

\usepackage{graphicx}	
\usepackage{amsmath}	
\usepackage{amssymb}	
\usepackage{multicol}        
\usepackage{comment}
\usepackage{bm}		
\usepackage{pdflscape}	
\usepackage{xcolor}

\def\nel{n_{{\rm e}^-}}
\def\np{n_{\rm p}}
\def\mel{m_{{\rm e}^-}}
\def\mp{m_{\rm p}}

\def\m2a{M^2_{\rm \small A}}

\def\vb{v_{\small 1}}

\def\vfs{V_{\rm \small FS}}

\def\vj{V_{\rm j}}
\def\vrs{V_{\rm \small RS}}
\def\vs{V_{\rm s}}
\def\lsim{\lower.5ex\hbox{$\; \buildrel < \over \sim \;$}}
\def\gsim{\lower.5ex\hbox{$\; \buildrel > \over \sim \;$}}
\newcommand{\RN}[1]{
}

\usepackage[T1]{fontenc}
\usepackage{ae,aecompl}

\title[]{Exact solution of one dimensional relativistic jet with relativistic equation of state}

\author[Joshi, Chattopadhyay, Ryu \& Yadav]
{Raj Kishor Joshi$^{1,3}$, Indranil Chattopadhyay$^{1}$\thanks{Email:
indra@aries.res.in (IC)}, Dongsu Ryu$^{2}$, Lallan Yadav$^{3}$\\
$^{1}$Aryabhatta Research Institute of Observational Sciences 
(ARIES), Manora Peak, Nainital-263002, India.\\
$^{2}$Department of Physics, School of Natural Science, UNIST, Ulsan 44919,
Korea.\\
$^{3}$Deen Dayal Upadhyaya Gorakhpur University, Gorakhpur, Uttar Pradesh-273009, India
}

\pubyear{2020}

\begin{document}
\label{firstpage}
\pagerange{\pageref{firstpage}--\pageref{lastpage}}
\maketitle

\begin{abstract}
We study the evolution of one-dimensional relativistic jets, using the exact solution of the Riemann problem for
relativistic flows. For this purpose, we solve equations for the ideal special relativistic fluid composed of dissimilar
particles in flat space-time and the thermodynamics of fluid is governed by a relativistic equation of state. We obtain
the exact solution of jets impinging on denser ambient media. The time variation of the cross-section of the jet-head is
modeled and incorporated. We present the initial condition that gives rise to a reverse shock. If the jet-head cross-section
increases in time, the jet propagation speed slows down significantly and the reverse-shock may recede opposite
to the propagation direction of the jet.
We show that the composition of jet and ambient medium
can affect the jet solution significantly. For instance, the propagation speed depends on the composition and is maximum for a
pair-dominated jet, rather than a pure electron-positron or electron-proton jet. The propagation direction of
the reverse-shock may also strongly depend on the composition of the jet. 
\end{abstract}



\begingroup
\let\clearpage\relax
\tableofcontents
\endgroup
\newpage

\section{Introduction}
\label{sec:intro}
Astrophysical jets were first discovered by Curtis \citep{c18} while studying the M87 galaxy in optical wavelength. 
With the advent of radio telescopes, the subject of astrophysical jets started to flourish. Now astrophysical jets are
studied at least in three wavebands, radio, optical, and X-rays. All these detailed observations revealed that astrophysical jets
are quite common and are associated with a variety of objects 
like AGNs (active galactic nuclei), microquasars, GRBs (gamma-ray bursts), and even YSOs (young stellar objects), albeit in a
vastly varied length scales and energetics.
Out of these, the jets associated with AGNs, microquasars, and GRBs are relativistic in terms of bulk speed as well as, temperature.
After analysing the observations, a general picture emerged that the jets originate from a region very close to the compact objects
like black holes residing at the heart of AGNs and microquasars \citep[for example in M87][]{jbl99,detal12}.
From the detailed observations of microquasars, the jet states were found to correlate with the accretion states, suggesting
that the jets originated from the accretion discs \citep{gfp03,fgr10,rsfp10}. 
However, even after the vast improvement in observational facilities, there is no consensus about the composition, formation, and
collimation of the jets.

Investigations of astrophysical jets, on one hand, involve the formation problem and on the other, the propagation problem.
In this paper, we are interested to study the propagation properties of jets.
One of the earliest models for the jet flow was proposed to be a version of the de Laval nozzle by \citet{br74}.
This paper outlined the basic jet structure of a supersonic jet beam far away from the central object. The jet 
after interaction with the ambient medium, develops a reverse shock in the jet beam, a contact discontinuity, and a forward shock.
The separation surface between the
jet and the ambient medium is called the contact discontinuity (CD). The forward shock (FS) is the shock transition in the ambient
medium as the jet head rams through the ambient medium. And the reverse shock (RS) is in the jet beam. RS and FS are on either side
of a CD. \citet{mg85} associated the flares in the astrophysical jets with traveling shocks along an adiabatic conical jet.

Relativistic jets are studied by solving the equations of relativistic hydrodynamics (RHD). Early attempts involved
using explicit finite difference code with monotonic transport \citep{w72}. However these kind of codes depended on artificial
viscosity techniques to capture shocks.
Studies of the interaction of the jet with the ambient medium and the evolution of jet morphology got a tremendous boost with the
advent of high resolution shock capturing (or, HRSC) numerical simulation methods \citep{vp93,mmi94,dh94,mmfi95,am03,wa13}. HRSC methods solve the fluid equations in the conserved form. The flow variables in the observer frame, also called
`state vectors', are updated in time by evaluating the fluxes of these state vectors at the cell surfaces.
These fluxes are computed by means of exact or approximate Riemann solvers. RHD codes based on exact Riemann solvers were
developed by \citet{mm96,wpl97}. There are great variety of numerical schemes available using approximate Riemann solvers.
\citet{b94} developed an approximate Riemann solver using the jump conditions in the shock frame while \citet{dw97} followed
a similar path but by using conditions for obliques shocks. \citet{e93,em95} extended the Roe-type approximate Riemann solver\citep{r81}
to RHD. Higher order reconstructions were also proposed \citep{mmimd92,dw95}. More detailed account on the development of
HRSC methods can be found in \citet{mm03, mm15}.

A global picture of the flow with multiple shocks within a jet has emerged from these simulations.

Most of the simulations were based on the ideal gas equation of state \citep[with a notable exception of][]{samgm02} which
is a reasonable approximation when the flow remains sub-relativistic or extremely relativistic, but the jets travel over a long
distance and the jet material can go through a transition from the relativistic to the non-relativistic regime and vice versa.
It may be noted that a flow can be called relativistic on the account when its bulk velocity $v\sim c$ (where $c$ is the speed of
light in vacuum)
or if the thermal energy of the flow $kT$ ($k$ is the Boltzmann constant and $T$ is the temperature) is of the order, or greater
than the rest mass energy of the gas particles. 
Assuming relativistic Maxwell-Boltzmann distribution of particles, the energy density of the flow or the equation of state (EoS)
of the fluid, was computed independently by
various authors \citep{c38,s57,cg68}, which were a combination of modified Bessel's function of various kinds.
This EoS is relativistically perfect and is abbreviated as RP. By using the recurrence relation, \citet{vkmc15} showed the
equivalence between the different forms of
the RP EoS, obtained by various authors mentioned above. One of the features of such relativistic EoS is that one need not specify
any adiabatic index to describe the
thermodynamics of the fluid. The adiabatic index is a function of temperature and can be automatically obtained if the temperature
is known.
Although it is possible to implement the RP in numerical simulation codes, but the presence of the modified Bessel function makes
it
computationally expensive \citep{fk96}.
Additionally, \citet{t48} in his seminal paper of relativistic shock adiabat, also obtained a relation between thermodynamic
variables of the
fluid in the form of a fundamental inequality which any EoS of relativistic fluid should obey, and is called the `Taub's inequality' or TI.
Any approximate EoS, apart from being a very good fit of RP, should simultaneously obey TI too.
To circumvent the problem of using RP at an extra computational cost, a number of approximate relativistic EoS were proposed
by various authors \citep{mpb05,rcc06}.
The EoS used by \citeauthor{mpb05} is actually the lower limit of TI.
On the other hand, \citet{rcc06} proposed a more accurate approximate EoS, and yet the energy density or the
enthalpy of the fluid is an algebraic function of pressure and mass density. One can compute the adiabatic index ($\Gamma$) from this EoS and it gave
correct asymptotic values at non-relativistic and relativistic temperatures. \citet{cr09}, extended the EoS of \citeauthor{rcc06}
for the fluids composed of dissimilar particles and the EoS is abbreviated as CR.
In this work, we use CR EoS which has been applied to a variety of astrophysical scenarios
\citep{cc11,cdmrs14,ck16,vc18,sc19,sc20,sic19,ddmc18}. 

An astrophysical jet is a relativistic beam plying through an ambient medium. The beam of the jet, because it is relativistic
should be less dense than the surrounding medium. Therefore,
the propagation of jets through a medium is essentially the time evolution of an initial discontinuity between
two states of a fluid which on one side is lighter and fast, and denser and static on the other side.
There might be a pressure/composition jump across the initial discontinuity or the pressure/composition
may also be uniform. Such an initial value problem is generally known as the Riemann problem. Depending on the physical condition of the initial discontinuity, a Riemann problem might evolve into a shock-tube problem,
oppositely moving shock waves or oppositely moving rarefaction waves, a wall shock, etc.
In the Newtonian regime, solutions of the Riemann problem played an important role in testing several hydrodynamic codes
\citep{s78}. Most of the modern hydrodynamic codes in the Newtonian regime are based on exact or approximate Riemann problem
solutions \citep{l92}, such that building a better Riemann solver (exact or approximate),
has become a very important field of research \citep{t97}.
Codes developed based on the above techniques and in the Newtonian regime, have been used extensively to study
diverse fields like the formation of stars
in the galactic plane \citep{ko01}, supernovae ejecta propagation \citep{afm89}, cosmological simulations \citep{rokc93},
accretion discs \citep{rbol95,rcm97}, etc.

In 1994, \citeauthor{mm94} derived analytical
solutions of the Riemann problem for special relativistic fluid, but only for
the flow with velocity component normal to the initial discontinuity. This paper helped to test and even develop various
numerical simulation codes for relativistic fluid
\citep[also see][]{lcgg13}. However, because
of the existence of the upper limit of the fluid velocity, the velocity components are not entirely independent of each other.
As a result,
the form of the eigenvalues of a flow with {normal and tangential velocity components with respect to the discontinuity, are
different from that of a flow
with only a normal velocity component}. \citet{pmm00} generalized the analytical Riemann problem of relativistic fluid obtained by
\citeauthor{mm94} for arbitrary tangential velocity components. 
The Riemann problem associated with relativistic jets is characterised by an FS,
a CD or the jet head, and the RS somewhere in the jet beam. The solution of Riemann
problem with a fixed adiabatic index EoS is relatively easier, but we will discuss later that such a solution is non-trivial
with a relativistic EoS like CR. Although, exact solutions of the Riemann problem are generally used to test simulation codes,
but these problems resemble astrophysical scenarios, so if judiciously used, exact solutions of the Riemann problem can be used
to study astrophysical problems as well \citep{hh19}.

In this paper, we solve the Riemann problem associated with a relativistic jet. The thermodynamics of the flow is described
by CR EoS. All the basic features of the relativistic jet are discussed considering an electron-proton flow.
In this paper, we would like to find the necessary condition for which the initial discontinuity develops into two shocks, the
reverse shock in the jet beam and a forward shock in the ambient medium ahead of the jet head.
We would like to investigate how the expanding jet-head cross-section affects the evolution of the jet. We also study the effect of jet fluid composition
on the overall jet evolution.
We investigate the effect of the composition of the jet beam on jet evolution, while keeping the ambient
composition same. We compared the effect
of the ambient medium composition on jet evolution, while keeping the jet composition same.
For the sake of completeness, we have given a short account of the solution
of a relativistic shock tube problem and wall-shock problem in the Appendix. We also
compared the exact solution of the wall-shock
with the relativistic TVD simulation code \citep{rcc06,crj13} in the Appendix.

In section \ref{eqns}, we present the governing equations. In section \ref{subsec:eos}, we describe the CR EoS for fluid
composed of dissimilar particles. In section \ref{sec:shock}, the structure of relativistic shocks is described.
We discuss the methodology to obtain the solution in section \ref{sec:meth}.
In section \ref{subsec:2sokcond}, we derive the condition for the formation of two shocks of a jet.
In section \ref{subsec:expndjet}, we discuss the effect of an expanding cross-section of the jet head, on the
structure of the jet. In sections \ref{subsec:compo}, \ref{subsec:samcompo}, \ref{subsec:diffcompo}, we discuss the effect of
fluid composition on the evolution of jets. And finally, in section \ref{sec:conclude} we discuss the highlights and summarize
the results. In addition, we also 
present exact solutions of two types of
relativistic Riemann problem, one is the shock-contact-rarefaction fan
problem or a shock-tube problem and two, shock-contact-shock 
with CR EoS in Appendix \ref{app:riem}.

\section{Governing equations of relativistic hydrodynamics}
\label{eqns}
We study ideal, relativistic fluid in flat space-time, and the energy-momentum tensor of such a fluid is given by;
\begin{equation}
  T^{\mu\nu}=\rho hu^{\mu}u^{\nu}+p\eta^{\mu\nu},
  \label{eq:tmunu}
\end{equation}
where $\rho$, $p$, and $h$ are the rest-mass density, local pressure, and the specific enthalpy of the fluid, respectively.
We follow the convention where the Greek indices represent space-time components of the vectors and tensors.
Contravariant components of the four-velocity are represented by $u^{\mu}$ and $\eta^{\mu\nu}$ are the components of Minkowski metric tensor in the Cartesian coordinates.
\begin{equation}
\eta^{\mu\nu}=\textrm{diag(-1,1,1,1)}
\label{eq:strmet}
\end{equation}  
Four-velocity of the fluid satisfies the normalization condition i.e.
\footnote{Throughout this paper we will use unit system where speed of light $c$ is set to unity, unless mentioned
otherwise} \\
\begin{equation}
u^{\mu}u_{\mu}=-1
\label{eq:forvel}
\end{equation}

The conservation of mass flux and energy-momentum gives us the relativistic fluid equations of motion.
\begin{equation}
(\rho u^{\nu}),_{\nu}=0
\label{eq:cont1}
\end{equation}
\begin{equation}
T^{\mu\nu},_{\nu}=0
\label{eq:momconserv}
\end{equation}
Using the normalization condition (\ref{eq:forvel}) we can write four-velocity as 
\begin{equation}
u^{\mu}=\gamma(1, v^x, v^y, v^z) ~~~[v_i=v^i \mbox{ components of three velocity}]
\label{eq:fourv}
\end{equation}
where $\gamma$ is the Lorentz factor 
\begin{equation}
\gamma=(1-v^2)^{-\frac{1}{2}}
\label{eq:lorenz}
\end{equation}
and  
\begin{equation}
v^2=(v^x)^2+(v^y)^2+(v^z)^2
\label{eq:thrivel}
\end{equation} 

In Minkowski space-time, the equations of relativistic hydrodynamics can be written in conservative form  
\begin{equation}
\partial_t\textbf{U}+\partial_i\textbf{F}^{(i)}=0 
\label{eq:eomconsv}
\end{equation}
Where \textbf{U} and $\textbf{F}^{(i)}$ (\textit{i}$\equiv$x, y, z) are the vectors and fluxes of the conserved variables,
respectively.
\begin{equation}
\textbf{U}=(D, M^x, M^y, M^z, E)^T
\label{eq:statvec}
\end{equation} 
\begin{equation}
\textbf{F}^{(i)}=(Dv^i, M^xv^i+p\delta^{xi}, M^yv^i+p\delta^{yi}, M^zv^i+p\delta^{zi}, (E+p)v^i)^T
\label{eq:flux}
\end{equation} 
where conserved variables $D$, $M^i$, and $E$ denote the mass density, momentum density, and energy density, 
respectively. These conserved variables can also be written in terms of primitive variables,   

\begin{equation}
\begin{array}{lll}
D =\rho\gamma \\
\\
M^i=\rho h \gamma^2 v^i \\
\\
E=\rho h\gamma^2-p
\end{array}
\label{eq:conv}
\end{equation} \\
The equation of state (EoS) is used to close the set of equations (\ref{eq:eomconsv}). The EoS can be written in form 
\begin{equation}
e=e(p,\rho)
\label{eq:eos}
\end{equation} 
where $e$ is the energy density in the local frame.

The set of equations of motion (\ref{eq:cont1}, \ref{eq:momconserv} or \ref{eq:eomconsv}) are hyperbolic in nature
and admits five real eigenvalues. Three of which are degenerate and are the entropy mode, while the first
and the last one are non-degenerate and are the acoustic modes. The eigenvalues are:
\begin{equation}
 \begin{array}{lll}
 \beta_1= \frac{v^x(1-c_s^2)- c_s\sqrt{(1-v^2)[1-v^2c_s^2-(v^x)^2(1-c_s^2)]}}{1-v^2c_s^2} \\
 \\
 \beta_2= v^x \\
 \\
 \beta_3= v^x \\
 \\
 \beta_4= v^x \\
 \\
 \beta_5=\frac{v^x(1-c_s^2)+ c_s\sqrt{(1-v^2)[1-v^2c_s^2-(v^x)^2(1-c_s^2)]}}{1-v^2c_s^2}
 \end{array}
\label{eq:eigenval}
\end{equation}
Here $c_s=\sqrt{(\partial p/\partial e)_s}$ is the adiabatic, relativistic sound speed.
The full eigenstructure with right and left eigenvectors for relativistic fluid has also been obtained
previously for a general EoS \citep[see][]{rcc06}.

\subsection{Equation of State (EoS)}
\label{subsec:eos}
We use CR EoS \citep{cr09} for the fluids composed of electrons, positrons, and protons.
The CR EoS is of the following form
\begin{equation}
e=\Sigma_i\left(n_im_ic^2+p_i\frac{9p_i+3n_im_ic^2}{3p_i+2n_im_ic^2}\right)
\label{eos_old.eq}
\end{equation}
In Eq. \ref{eos_old.eq}, $c$ is used explicitly.
The index $i$ represents the various species that constitute the fluid and $c$ is the speed of light in
vacuum. In this paper, we consider
the fluid to be composed of electrons, protons, and positrons of various proportion.
The above equation can be represented in the unit system where $c=1$, as, 
\begin{equation}
 e=\rho f,
\label{eq:eos2}
\end{equation}
where,
\begin{equation}
f=1+(2-\xi)\Theta\left[\frac{9\Theta+6/\tau}{6\Theta+8/\tau}\right]+\xi\Theta\left[\frac{9\Theta+6/\eta\tau}{6\Theta+8/\eta\tau}\right]
\label{eq:eos3}
\end{equation}
In the above equations $\rho=\Sigma_i n_im_i=\nel \mel (2-\xi+\xi/\eta)$, where $\xi=\np/\nel$, $\eta=\mel/\mp$ and
$\nel$, $\np$, $\mel$ and $\mp$ are the electron number
density, the proton number density, the electron rest mass, and proton rest mass.
Moreover, the ratio of the pressure and the local rest energy density of fluid, is a measure of temperature
$\Theta=p/\rho$ and $\tau=2-\xi+\xi/\eta$ \footnote{Although the EoS in equation \ref{eq:eos2} is exactly
same as that in \citet{cr09}, but in this paper, $\tau$ is included in the definition
of $\Theta$ and $f$}. \\
The specific enthalpy is given as  
\begin{equation}
h=(e+p)/\rho=f+\Theta;
\label{eq:enthalp}
\end{equation}
The expression for polytropic index N is given as 
\begin{eqnarray}
&N =  \rho \frac{\partial h}{\partial p}-1=\frac{\partial f}{\partial \Theta}=6
\left[(2-\xi)\frac{9\Theta^2+24\Theta/\tau+8/\tau^2}{(6\Theta+8/\tau)^2}\right]\\ \nonumber
& +6\xi \left[\frac{9\Theta^2+24\Theta/(\eta \tau)+8/(\eta \tau)^2}{\{6\Theta+8/(\tau \eta)\}^2}\right]
\label{eq:poly}
\end{eqnarray}

\begin{figure}
\hspace{0.0cm}
\includegraphics[width=\columnwidth]{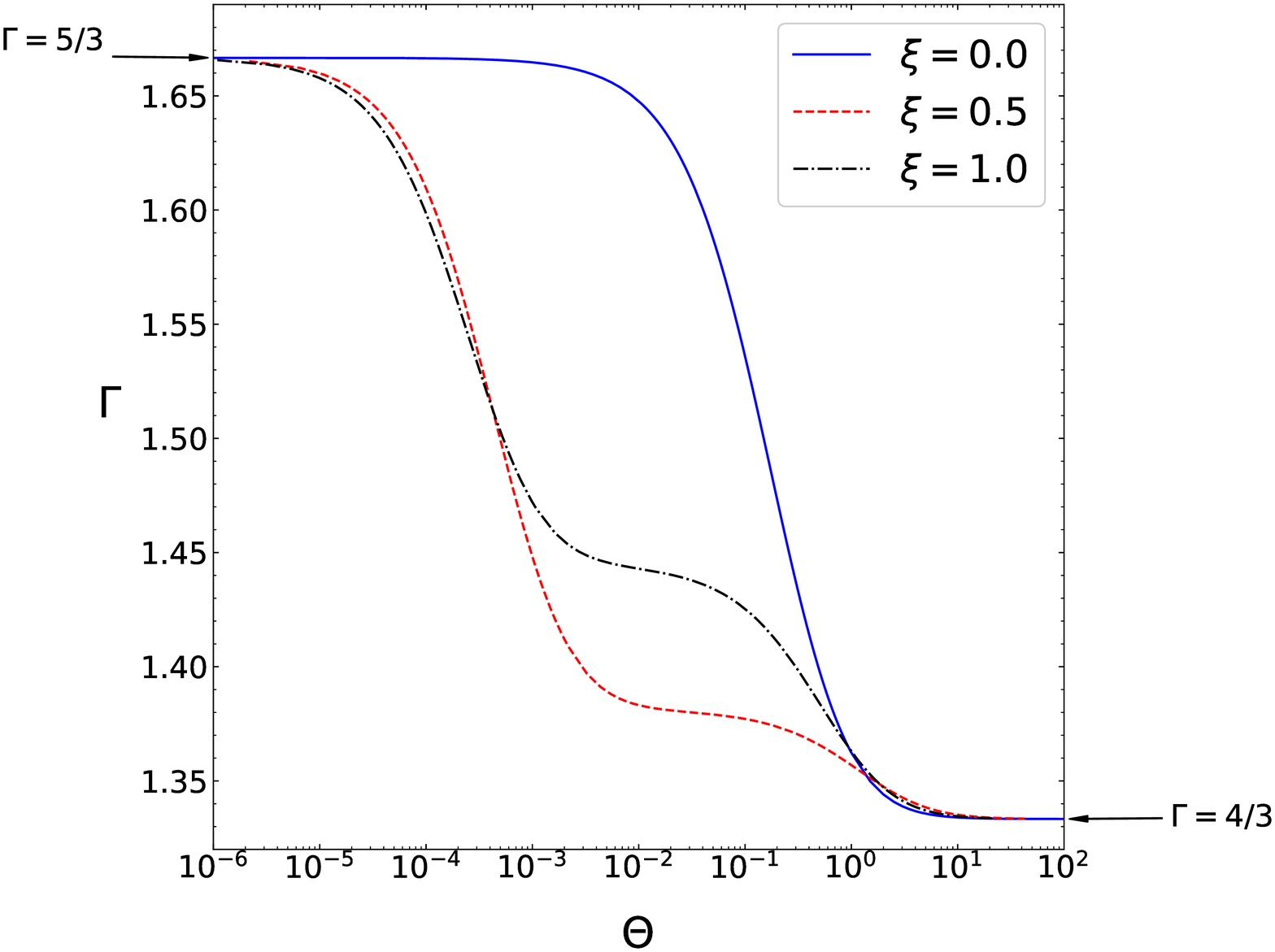}
\caption{Variation of adiabatic index with $\Theta$ for fluids with composition parameter
$\xi=0.0~\mbox{(solid blue)},~\xi=0.5~\mbox{(red dotted)},~\xi=1.0~\mbox{(black dot-dashed)}$.}
\label{fig:gamma_cr_eos}
\end{figure}

The polytropic index is not a constant but a function of $\Theta$ and $\xi$.  It may be noted that,
$N\rightarrow 3$ as $\Theta \gg 1$; while $N\rightarrow 3/2$ as $\Theta \ll 1$, therefore $N$ approaches asymptotic
values at very high and low temperatures.
It may be noted further that, for $\xi=0$ (i. e., single species gas), the expression of the polytropic index is exactly
same as that was presented in \citet{rcc06}. And the adiabatic index $\Gamma$ is  
\begin{equation}
\Gamma=1+\frac{1}{N}
\label{eq:adiab}
\end{equation}
The variation of adiabatic index $\Gamma$ with respect to $\Theta$ is shown in Fig. (\ref{fig:gamma_cr_eos}).
The composition of the fluid is marked in the legend, electron-positron or $\xi=0$ (solid, blue), equal proportion of positrons
and protons or $\xi=0.5$ (dashed, red) and the electron-proton fluid or $\xi=1.0$ (dashed-dot, black).
It is quite clear that $\Theta>10$ is
the ultra-relativistic temperature regime i. e. $\Gamma \rightarrow 4/3$, for a fluid of any composition $\xi$.
On the other hand, $\Theta<10^{-5}$ is the non-relativistic temperature regime, where $\Gamma \sim 5/3$ for fluid
with $\xi>0$. But for $\xi=0$, $\Gamma \rightarrow 5/3$ is for $\Theta \lsim$ few$\times 10^{-3}$.
Clearly, the thermodynamics of the flow depends on both the temperature and the composition $\xi$ of the flow.
It may be remembered that the conversion between the absolute temperature $T$ and $\Theta$ is given by
$T={\tau \mel \Theta}/{2k}$.  
The expression of sound speed starting from the first principle can be written as,
\begin{equation}
c_s^2=\frac{1}{h}\frac{\partial p}{\partial \rho}=-\frac{\rho}{Nh}\frac{\partial h}{\partial \rho}=\frac{\Gamma\Theta}{h}
\label{eq:soundsp}
\end{equation}
For non-relativistic temperatures or $\Theta \ll 1$, $h \rightarrow 1$ and $\Gamma \rightarrow 5/3$,
sound speed approaches non-relativistic value $c_s \rightarrow \sqrt{5\Theta/3}$. In contrast, for $\Theta \gg 1$,
$h \sim 4 \Theta$, $\Gamma \sim 4/3$, so $c_s \rightarrow 
1/\sqrt{3}$, once again the sound speed achieves asymptotic values for ultra-relativistic and non-relativistic temperature
limit.

\subsection{Relativistic shock waves }
\label{sec:shock}
The information about the states on both sides of a shock is obtained by the jump conditions based on the continuity of mass and
energy-momentum fluxes. These conditions are known as the Rankine-Hugoniot (RH) conditions. The relativistic version of RH
conditions was obtained by \cite{t48}. The relativistic RH conditions are given as \citep[also see][]{pmm00}
\begin{equation}
[\rho u^\mu]n_\mu=0
\label{eq:shock1}
\end{equation}
\begin{equation}
[T^{\mu\nu}]n_\nu=0
\label{eq:shock2}
\end{equation}

where $n_\mu$ is a unitary normal vector to shock surface $\Sigma$ and we have used the notation.
Here,
\begin{equation}
[F]=F_a-F_b
\label{eq:sokdif}
\end{equation}\\
$F_a$ and $F_b$ are the values of the function $F$ on the either sides of the shock surface $\Sigma$.
Considering $\Sigma$ to be normal to the x axis and using the unitarity of $n_\mu$,  we can write it as
\begin{equation}
n^\mu=\gamma_s(\vs,1,0,0),
\label{eq:soknorm}
\end{equation}\\
where $\vs$ is the speed of shock (the speed of surface $\Sigma$) and $\gamma_s$ is the Lorentz factor of the shock. 
\begin{equation}
\gamma_s=\frac{1}{\sqrt{1-{\vs}^2}}
\label{eq:lorenzshok}
\end{equation}\\
We can introduce the invariant relativistic mass flux $j$ across the shock as
\begin{equation}
j=\gamma_sD_a(\vs-v_a^x)=\gamma_sD_b(\vs-v_b^x)
\label{eq:masflux}
\end{equation}
In the above the suffix `a' and `b' denoted the quantites on either side fo the shock.
The positive (negative) value of $j$ represents the shock propagating to right (left).
Multiplying equation (\ref{eq:shock2}) with $n_\mu$ and using the definition of $j$ (equation \ref{eq:masflux}),
we obtain the expression for relativistic mass flux in terms of pressure, enthalpy and density
\begin{equation}
j^2=-\frac{[p]}{[h/\rho]}
\label{eq:masflux2}
\end{equation}
We can write the RH conditions equations (\ref{eq:shock1}) and (\ref{eq:shock2}) in terms of conserved quantities $D,~M^j,$ and $E$ as   
\begin{equation}
[v^x]=-\frac{j}{\gamma_s}\left[\frac{1}{D}\right]
\label{eq:vxjump}
\end{equation}
\begin{equation}
[p]=\frac{j}{\gamma_s}\left[\frac{M^x}{D}\right]
\label{eq:presjump}
\end{equation}
\begin{equation}
\left[\frac{M^y}{D}\right]=0
\label{eq:myjump}
\end{equation}
\begin{equation}
\left[\frac{M^z}{D}\right]=0
\label{eq:mzjump}
\end{equation}
\begin{equation}
[v^xp]=\frac{j}{\gamma_s}\left[\frac{E}{D}\right]
\label{eq:vxp}
\end{equation}
Equations (\ref{eq:myjump}) and (\ref{eq:mzjump}) imply 
\begin{equation}
h\gamma v^{y,z}=\textrm{constant}
\label{eq:transvel}
\end{equation}
\begin{equation}
v_b^{y,z}=h_a\gamma_av_a^{y,z}\left[\frac{1-(v_b^x)^2}{h_b^2+(h_a\gamma_av_a^t)^2} \right]^{1/2}
\label{eq:transvel2}
\end{equation}
Where $v^t$ is the absolute value of tangential velocity of the flow. 
\begin{equation}
v^t=\sqrt{(v^y)^2+(v^z)^2}
\end{equation}
The expression for normal flow velocity $v^x$ is obtained by using equations (\ref{eq:vxjump}), (\ref{eq:presjump}) and
(\ref{eq:vxp}) 
\begin{equation}
v_b^{x}=\frac{\left(h_a\gamma_av_a^x+\frac{\gamma_s(p_b-p_a)}{j} \right)}{\left[ h_a\gamma_a+(p_b-p_a)
\left( \frac{\gamma_sv_a^x}{j}+\frac{1}{\rho_a\gamma_a}\right)\right]}
\label{eq:vbx}
\end{equation}
The expression for the shock velocity is obtained using the definition of mass flux  
\begin{equation}
{\vs}^{\pm}=\frac{\rho_a^2\gamma_a^2v_a^x\pm|j|\sqrt{j^2+\rho_a^2\gamma_a^2(1-(v_a^x)^2)}}{\rho_a^2\gamma_a^2+j^2}
\label{eq:shockvel}
\end{equation}
${\vs}^+\,\,({\vs}^-)$ corresponds to shock propagating towards right (left). \\

Multiplying equation (\ref{eq:shock2}) first by $(hu_\mu)_a$ and then by $(hu_\mu)_b$ and adding both the expressions results in
\begin{equation}
[h^2]=\left(\frac{h_b}{\rho_b}+\frac{h_a}{\rho_a}\right)[p]
\label{eq:taubadiabat}
\end{equation}  
            
Equation (\ref{eq:taubadiabat}) is known as Taub adiabat. For the general case Taub adiabat is solved along with the EoS to
obtain $h_b$ as a function of $p_b$.  

\begin{figure}
\includegraphics[width=\columnwidth]{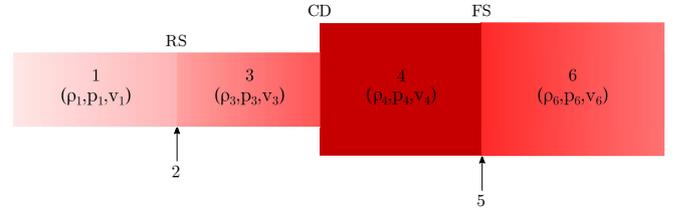} 
\caption{Cartoon diagram of the one dimensional jet (1DJ) structure. The jet is divided into six regions, where
region 1 is the original beam of the jet, and region 6 is the original ambient medium. Region 2 is the location of reverse shock (RS),
and region 5 is the location of the forward shock (FS). 
The CD is the location of the jet head (JH). The region 3 is the shocked jet region and region 4 is the shocked ambient medium.
Generally the cross-section may vary across every region.}
\label{fig:cartun}
\end{figure}

\begin{figure}
\includegraphics[width=\columnwidth]{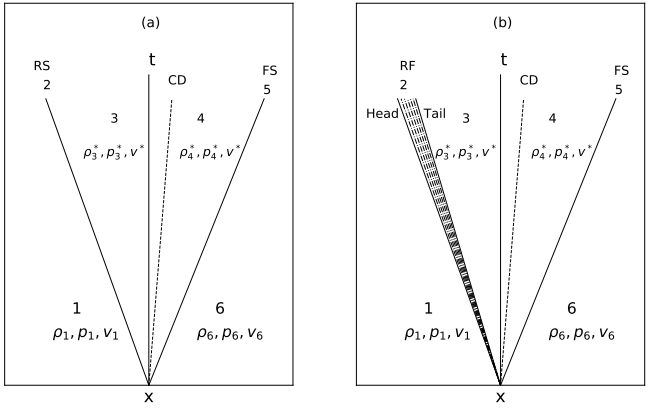} 
\caption{Schematic diagram for: (a) relativistic jet (1DJ) --- At $t=0$, region 1 is the beam of the jet, and region 6 is the ambient medium. The initial discontinuity evolves into a forward shock (FS), and contact discontinuity
(CD) and a reverse shock (RS). (b) Shock tube problem --- Region 1 and 6 are the initial left and right state and are marked as suffix of the flow variables in the respective regions. The initial discontinuity evolves into a CD flanked by an FS and an RF.}
\label{fig:schem}
\end{figure}

\section{Methodology}
\label{sec:meth}
A jet is characterized by low density, high velocity (high $v^x$), `geometrically narrow' material travelling with relativistic
velocity
through a denser, colder, static medium. We consider the relativistic jet as the time evolution of a one dimensional,
initial discontinuity. The defining distinction
being that the initial left state of the fluid is considered as the jet beam, which is lighter than the ambient
medium, moving with relativistic speed to the right. The surging jet beam drives a shock on the ambient medium
which also moves in the direction of propagation and is called the forward shock (FS). The initial discontinuity
evolves with local jet speed and is the jet-head (JH) or contact discontinuity (CD). The CD moves slower than the jet beam
and drives a shock in the jet beam and is called the reverse shock (RS).\\
Therefore, one dimensional jet (1DJ) consists of a forward and a reverse shock
separated by a CD.
The solution is represented as
six regions ({see Fig. \ref{fig:cartun}}), which are: 
\begin{itemize}
 \item Region 1: It is the initial unperturbed state which is characterised by injection parameters of jet beam,
 $\rho_1$, $p_1$ and $\vb$ (i. e., $v^x_1$).
 \item Region 2: is a thin surface of RS, which may move in the same direction as FS, but depending on the physical condition may also move
 in the opposite direction.
 \item Region 3: It is the region between the RS and the CD. The state is represented as 
$\rho_3^*,~p_3^*$ and $v_3^*$. 
\item Region 4: It is the region between CD and FS. There is a density jump across CD and the flow variables in this
region are represented as $\rho^*_4,~p_4^*$ and $v_4^*$.
\item Region 5 is the surface of FS travelling to the right.
\item Region 6: This region corresponds to the ambient medium which has not been influenced by the forward shock wave and
the flow variables in this region are represented as $\rho_6,~p_6,~\&~v_6$.
\end{itemize}
It may be noted that regions 2 and 5 are thin shock surfaces. However, depending on the initial conditions both these
surfaces may become rarefaction fan with non-negligible spatial extent (see, Figs. \ref{fig:schem}a,b for qualitative argument
and Figs. \ref{fig:features}a-d for a quantitative understanding). Therefore as a part of general approach,
these surfaces are called regions.
In Fig. (\ref{fig:schem}a), we present the schematic diagram of the time evolution of these surfaces in 1DJ. 
Figure (\ref{fig:schem}a) shows that the location of CD, FS, and RS at any time will be on
the three straight lines and the values of the flow variables in the six regions will be as mentioned above.
It is clear that the time evolution of 1DJ is similar to an initial value problem which is also known as the Riemann problem.
Riemann problems are of various kinds, one of the most popular example of one, is the so-called shock-tube (ST) problem.
In Fig. (\ref{fig:schem}b), we present the schematic diagram of an ST problem for comparison.
It may be remembered that ST too is an initial value problem, however, the initial conditions are such (high pressure/density)
that FS and CD forms but instead of an RS, a rarefaction fan (RF) is formed. The RF moves in a direction opposite to FS and CD. The detailed solution procedure for a couple of other Riemann problems is presented in the Appendix.
In this paper, we present the solution for 1DJ problem.\\

The time evolution and solution of the 1DJ problem, depends crucially on the measurement
of the post shock velocity. Recalling equation (\ref{eq:vbx}) we obtain an expression for $(v_3^x)^*$ in terms of $v_1^x$ for a RS and an expression for $(v_4^x)^*$ in terms of $v_6^x$ for a FS.
\begin{equation}
\begin{split}
&(v_3^x)^*=\left(h_1\gamma_1v_1^x+\frac{\gamma_s(p_*-p_1)}{j} \right)\\
&~~~~~~~\left( h_1\gamma_1+(p_*-p_1)\left( \frac{\gamma_sv_1^x}{j}+\frac{1}{\rho_1\gamma_1}\right)\right)^{-1}
\end{split}
\label{eq:v3jet}
\end{equation}
\begin{equation}
\begin{split}
&(v_4^x)^*=\left(h_6\gamma_6v_6^x+\frac{\gamma_s(p_*-p_6)}{j} \right)\\
&~~~~~~~\left( h_6\gamma_6+(p_*-p_6)\left( \frac{\gamma_sv_6^x}{j}+\frac{1}{\rho_6\gamma_6}\right)\right)^{-1}
\end{split}
\label{eq:v4jet}
\end{equation}
For equation (\ref{eq:v3jet}) $j$ is the negative root of equation (\ref{eq:masflux2}) and for equation (\ref{eq:v4jet}) it is the positive root
of equation (\ref{eq:masflux2}). The continuity of velocity across JH is the key to obtain the solution of this problem
\begin{equation}
(v_4^x)^*-(v_3^x)^*=0
\label{eq:velbalnc_jet}
\end{equation}
Equation (\ref{eq:velbalnc_jet}) is solved for $p^*$ using the iterative root finder and rest of the quantities can be calculated
once $p^*$ is obtained. Densities in Region 3 and 4 are obtained from the Taub adiabat for RS and right FS respectively. The $(v_3^x)^*=(v_4^x)^*=(v^x)^*$ is the speed $\vj$ with which the jet-head or CD is propagating.\\

x{The continuity of the normal component of velocity follows from the fact that the mass flux across the CD is zero}. 
\begin{equation}
j=\gamma_jD_a(\vj-v_a^x)=\gamma_{j}D_b(\vj-v_b^x)=0
\label{eq:jcd}
\end{equation}
Where $\vj$ is the velocity of jet head or the CD and $\gamma_j$ is the Lorentz factor of jet head.

\subsection{Expansion across Jet head}
{As the jet expands it can result in discontinuous cross sectional area across the jet head. To obtain the equations that govern
the dynamics of the flow when the area across CD/JH is not same, we use the momentum flux balance across the CD/JH \citep{myt04}
\begin{equation}
A_b\left[\rho_bh_b\gamma_j\gamma_b^2(v_b-\vj)^2+p_b\right]=A_a\left[\rho_ah_a\gamma_j\gamma_a^2(v_a-\vj)^2+p_a\right]
\label{eq:var_area_mb}
\end{equation}
Using equation (\ref{eq:jcd}) with (\ref{eq:var_area_mb}) we obtain 
\begin{equation}
A_bp_b=A_ap_a
\label{eq:pbal_var_area}
\end{equation}
And the velocity balance condition across the jet head is given as    
\begin{equation}
v_3^*(p_3^*)-v_4^*(p_4^*)=0 
\label{eq:vbal_expjet1}
\end{equation}
Assuming $p_4^*=p^*$, equation (\ref{eq:vbal_expjet1}) is a transcendental equation for variable  
$p^*$, as $p_3^*$ and $p_4^*$ are related by equation (\ref{eq:pbal_var_area}) as $A_3p_3^*=A_4p_4^*$.}\\

The jet kinetic luminosity is related with the jet cross-section
\begin{equation}
 L_j=\gamma_1^2(e+p)_1\vb \pi y_1^2,
 \label{eq:jetlum}
\end{equation}
Where quantities with suffix `1' are the jet beam variables where the jet beam velocity $\vb=v_1^x$, and $y_1$ is the
cross-sectional dimension of the jet beam, so $A_1=\pi y_1^2$.

\section{Results}
\label{sec:results}
\subsection{Formation of two shock fronts in relativistic jet}
\label{subsec:2sokcond}
The essential condition for the formation of two shock fronts is that the pressure in the
intermediate state $p^*$
(refer to Fig \ref{fig:schem}) should be greater than the pressure in the initial states.
\cite{rz01} have shown that the
intermediate pressure is a function of relative velocities between the initial states
so there will be a limiting
value of relative velocity for the occurrence of two shock fronts.
In this case, we assume the cross-sectional area of the flow to be invariant
across all the regions. To obtain the value of limiting
velocity we start by assuming the
scenario with a left and right shock separated by contact discontinuity. We assume the pressure
of jet beam $p_1$ is greater than the
pressure of ambient medium $p_6$.
\begin{figure*}
\hspace{0.0cm}
\includegraphics[width=14cm,height=10cm]{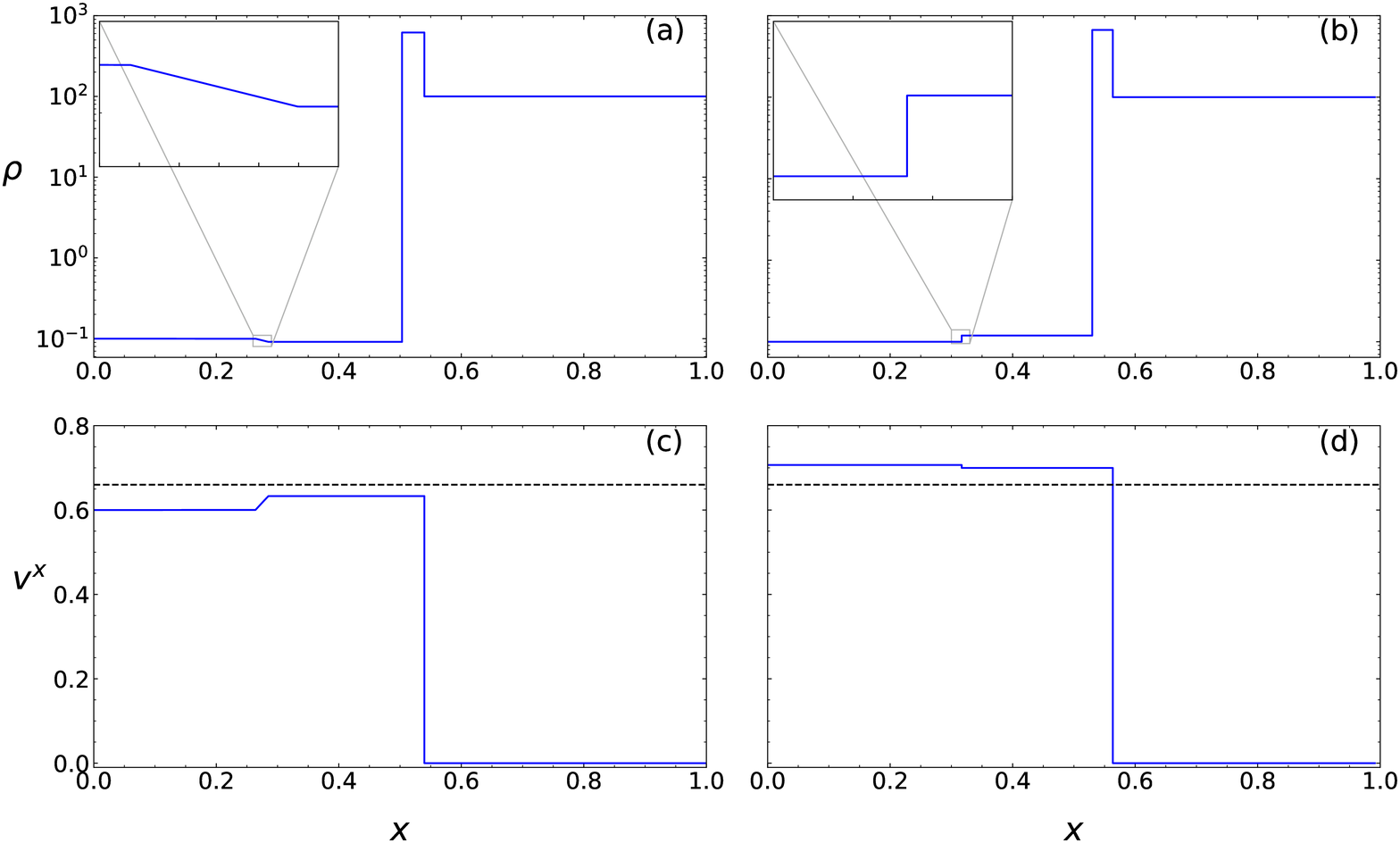}
\caption{Comparison of a shock-tube (a, c) and 1DJ solution (b, d). The injection velocities are $v_1=0.6$ (a, c) and
$v_1=0.7$ (b, d). Other parameters are $p_1=100,~\rho_1=0.1,~p_6=1.0,~\rho_6=100$ and the location of the initial
discontinuity is at $x_0=0.25$. The plot is at time $t=0.4$.
The composition parameter of injected beam and ambient medium is taken to be $\xi=1.0$.}
\label{fig:features}
\end{figure*}
Assuming there is an RS, the relative velocity of the pre-shock region (1) and the post-shock region (3)
is given by  \citep{rz01}
\begin{equation}
v_{13}=\sqrt{\frac{(p^*-p_1)(e_3-e_1)}{(e_1+p^*)(e_3+p_1)}}
\label{eq:relval1}
\end{equation}
Similarly, the relative velocity (between region 4 and 6) ahead of FS is given as 
\begin{equation}
v_{64}=-\sqrt{\frac{(p^*-p_6)(e_4-e_6)}{(e_6+p^*)(e_4+p_6)}}
\label{eq:relval2}
\end{equation}
It may be remembered, $v_3=v_4=v^*$ across the CD.
Hence the relative velocity of between two initial states is given as  
\begin{equation}
(v_{16})_{2S}=\frac{v_{13}-v_{64}}{1-v_{13}v_{64}}
\end{equation} 
For the limiting case where there exist an RS, the minimum value of pressure in the intermediate region
i.e. $p^*$, is given as   
\begin{equation}
p^*=\rm{max}(p_1, p_6)=p_1
\end{equation}
Hence the limiting value of relative velocity is given as 
\begin{equation}
\boldmath{v_{\rm lim}}=\sqrt{\frac{(p_1-p_6)(e_4-e_6)}{(e_6+p_1)(e_4+p_1)}}
\label{eq:critical_v}
\end{equation}
As the ambient medium is at rest, therefore the relative velocity between the initial states is
$(v_{16})_{2S}=({v_1-v_6})/({1-v_1 v_6})=v_1$.
\label{subsubsec:presmatch}
\begin{figure*}
\hspace{0.0cm}
\includegraphics[width=14cm,height=10cm]{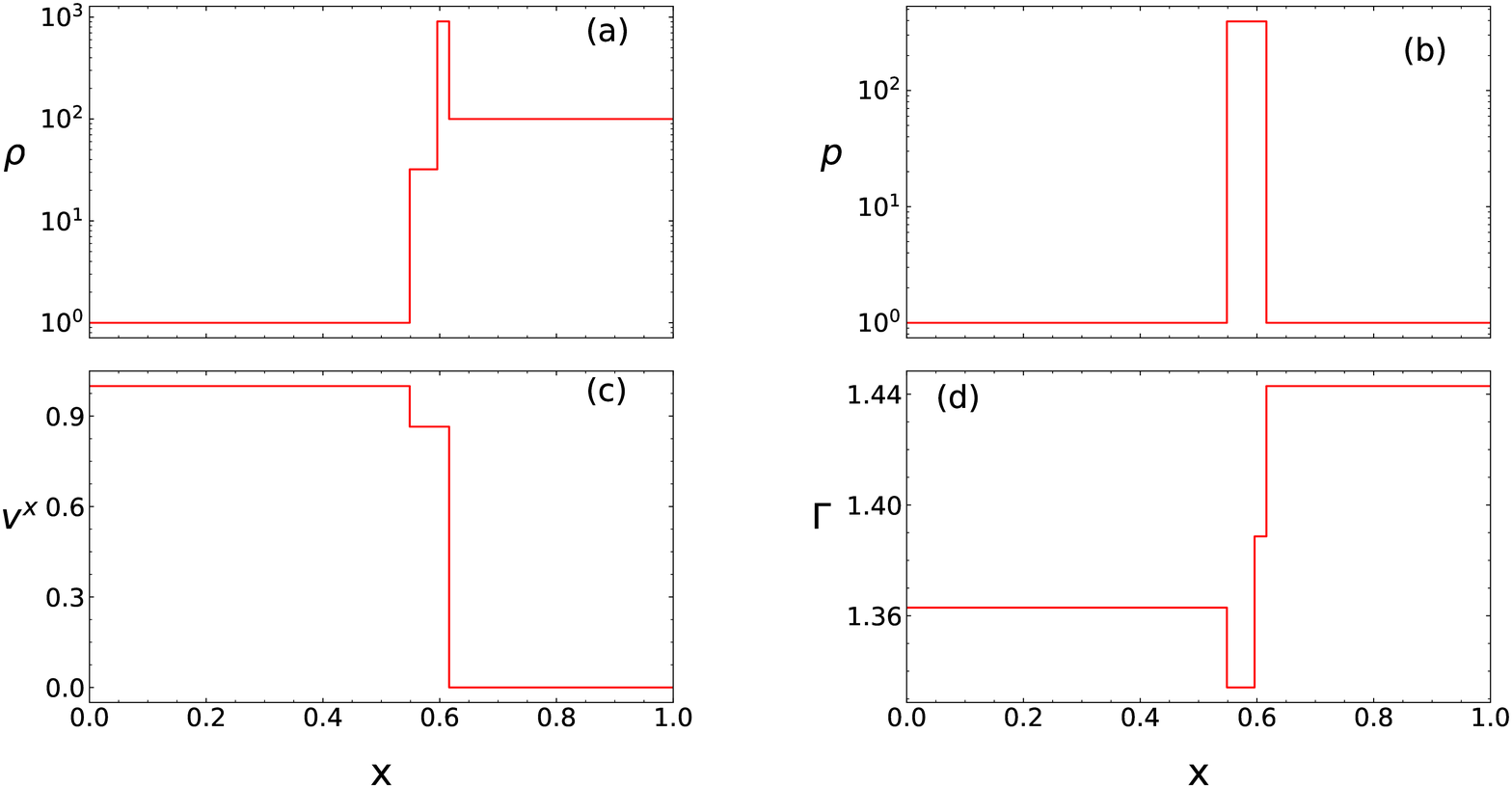}
\caption{Flow variables (a) $\rho$,(b) $p$, (c) $v^x$ and (d) $\Gamma$ as functions of $x$ at $t=1.0$ and the
fluid composition $\xi=1.0$. Initial conditions: $\rho_1=1.0,~p_1=1.0,~v_1=0.99944$ and $\rho_6=100.0,~p_6=1.0,~v_6=0.0$
and the initial discontinuity is at $x_0=0.001$.}
\label{fig:1dj}
\end{figure*}
For a particular density and pressure configuration, if the injection velocity $v_1$ is greater than the limiting velocity
$v_{\rm lim}$ (see, equation \ref{eq:critical_v}), then the injected jet beam evolves with an FS and an RS, otherwise, it evolves
as a standard shock-tube problem with an FS and an RF separated by CD. 
In Fig. (\ref{fig:features}a-d) we plot flows with two injection velocities, one with $v_1<v_{\rm lim}$ (Fig. \ref{fig:features}
a, c) and two with $v_1>v_{\rm lim}$ (Fig. \ref{fig:features}b, d). The initial parameters are given by 
\begin{equation}
p_1=100,~\rho_1=0.1,~p_6=1.0~\rho_6=100,~\mbox{and}~v_6=0
\label{eq:param}
\end{equation}
The composition parameter of beam and ambient medium is same ($\xi=1.0$). For the initial parameters (Eq. \ref{eq:param}),
the threshold injection velocity for the formation of
two shock front obtained from Eq. (\ref{eq:critical_v}) is $v_{\rm lim}=0.66$. The threshold level of injection velocity
is represented by the dashed line in Figs. (\ref{fig:features}c \& d).
Figure (\ref{fig:features}a \& c) show the variation of density and velocity, respectively, as functions
of position with beam injection velocity $v_1=0.60$. In Figs. (\ref{fig:features} b \& d) the variation of
$\rho$ and $v^x$ are for the case $v_1=0.70$.
It is clear that the initial discontinuity with the injection $v_1 < v_{\rm lim}$, evolves into a shock tube problem
with an CD flanked by an FS ahead and an RF behind it (Fig. \ref{fig:features}a \& c). The RF is zoomed
in Fig. (\ref{fig:features}a) for clarity.
The CD and FS going to the right while RF to the left.
On the other hand, in Fig. (\ref{fig:features}b \& d) $v_1 > v_{\rm lim}$, the initial discontinuity evolves into a solution
which has a CD flanked by an FS and RS. The RS is zoomed in the inset of Fig. (\ref{fig:features}b).

\subsubsection{Pressure matched jet}

For a pressure matched jet ($p_1=p_6$) the threshold value of relative velocity from equation (\ref{eq:critical_v}) is
obtained as
\begin{equation}
v_{\rm lim}=0
\label{eq:pmatch_vcrit}
\end{equation}
\begin{figure*}
\hspace{0.0cm}
\includegraphics[width=12cm,height=6cm]{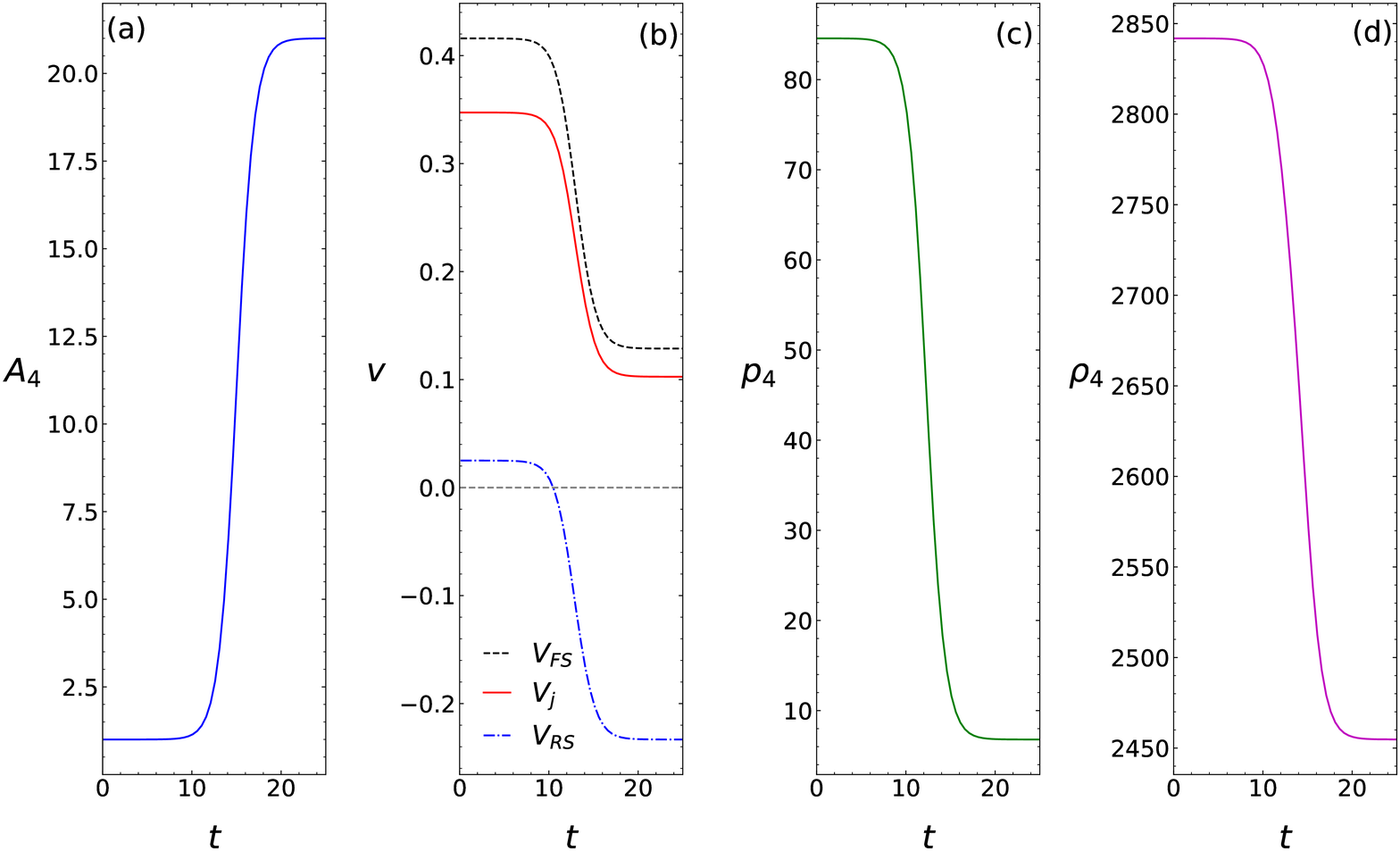}
\caption{(a) Variation of area ratio across the jet head with respect to time. (b) Variation of the RS, JH and FS velocities as
functions of time for an expanding electron-proton jet with injected Lorentz
factor $10$ and $p_1=p_6=0.1,~\rho_1=1.0,~\rho_6=500.0$ and $x_0=0.25$. The variation of pressure and density in region 4 which is the post shock
region of FS is shown in panels (c) and (d) respectively. }
\label{fig:profiles_jet}
\end{figure*}
Hence for any non zero injection velocity $v_1$, the pressure matched jet will always evolve as two shock fronts separated by a
contact discontinuity.  
Figure (\ref{fig:1dj}) shows the flow variables for a jet beam with initial parameters $\rho_1=1.0,~p_1=1.0,~v_1=0.99944$
injected into an ambient medium with $\rho_6=100.0,~p_6=1.0,~v_6=0.0$. We have assumed the uniform cross sectional flow for this
case.

The density jumps in Fig. (\ref{fig:1dj}a), show the positions of RS, CD, and FS (from left
to right, respectively). The shocked region of the jet beam is in between CD and RS and the kinetic energy of the jet
beam is converted to thermal energy in the post-shock region of RS. This shock heated region contains the most relativistic gas
($\Gamma\sim 4/3$). The shocked ambient medium is in between CD and FS. 
\begin{figure*}
\hspace{0.0cm}
\includegraphics[width=14cm,height=9cm]{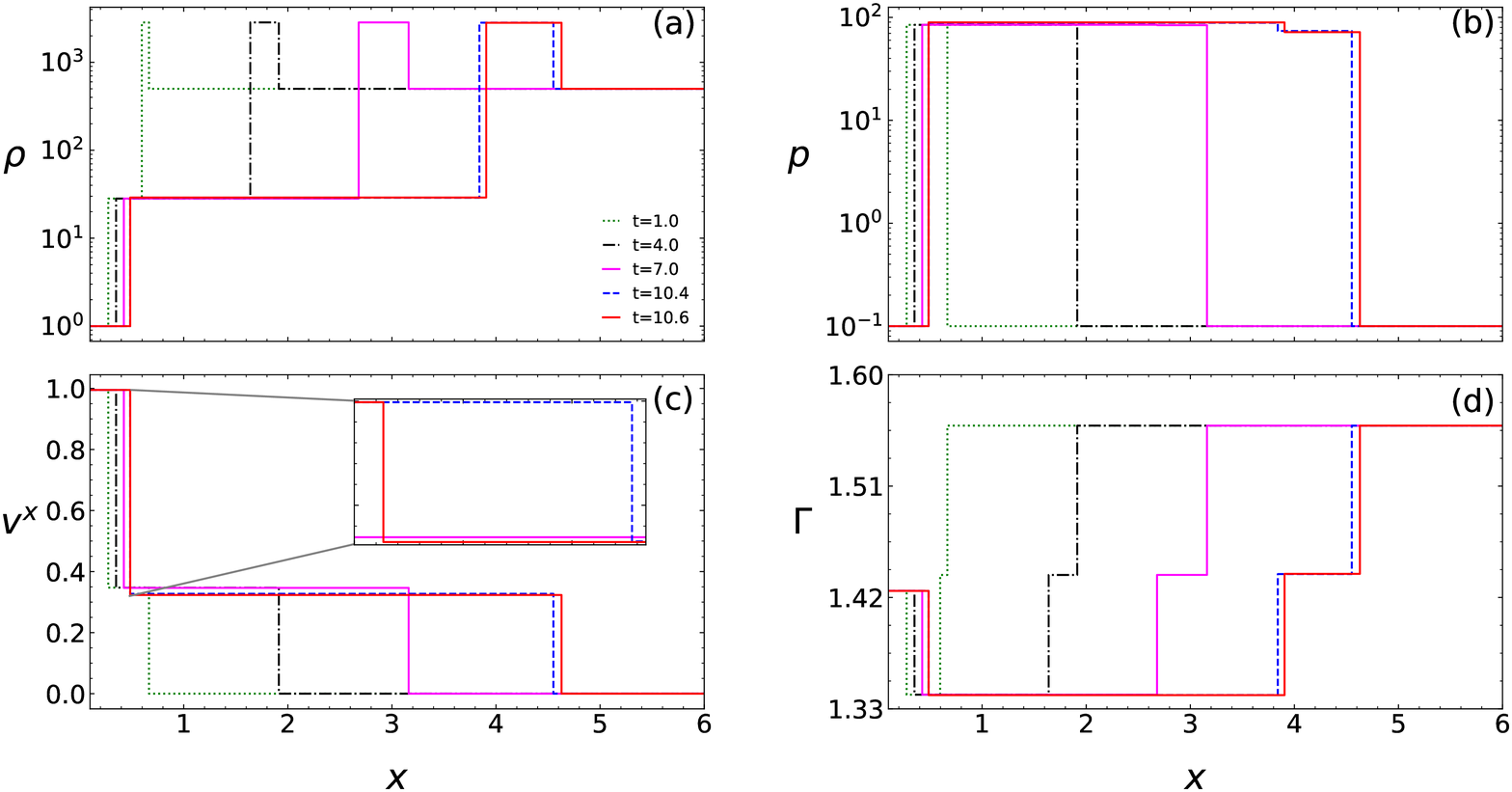}
\caption{Flow variables $\rho(a),~p(b),~v^x(c)\mbox{ and } \Gamma (d) $ as functions of $x$  at different time for an
expanding electron-proton jet. The jet solution are for the same parameters as
	in Figs. (\ref{fig:profiles_jet}).}
\label{fig:expanding_jet}
\end{figure*}

\begin{figure*}
\hspace{0.0cm}
\includegraphics[width=13cm,height=8cm]{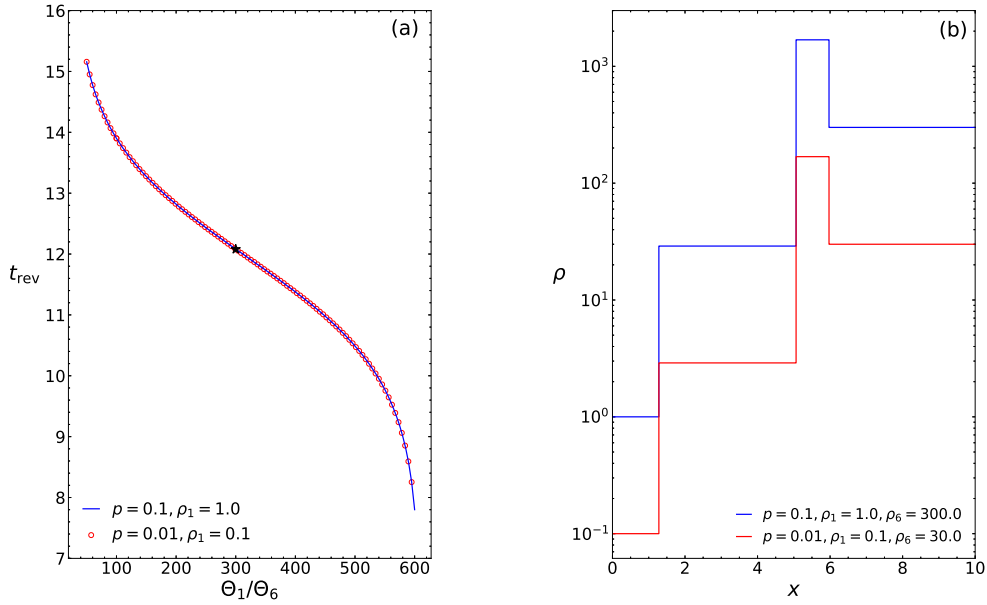}
\caption{(a) We plot the time $t_{\rm rev}$ taken by RS to move in the opposite direction of JH or CD as a function of
$\Theta_1/\Theta_6$. (b) The density profile $\rho$ w.r.t $x$ of the two jets marked by $\star$ ($\equiv t=12.08$) in panel (a).
The injection speed of both the electron-proton ($\xi=1$) jets $\vb=0.995$, but $p_1=p_6=p=0.1$ \& $\rho_1=1$ for one jet model (solid, blue) and
$p_1=p_6=p=0.01$ \& $\rho_1=0.1$ for the second jet model (open circle, red). $\rho_6$ for the jets, scale according
to the ratio $\Theta_1/\Theta_6$.}
\label{fig:revRSexpJ}
\end{figure*}
\begin{figure}
\hspace{0.0cm}
\includegraphics[width=7cm,height=6cm]{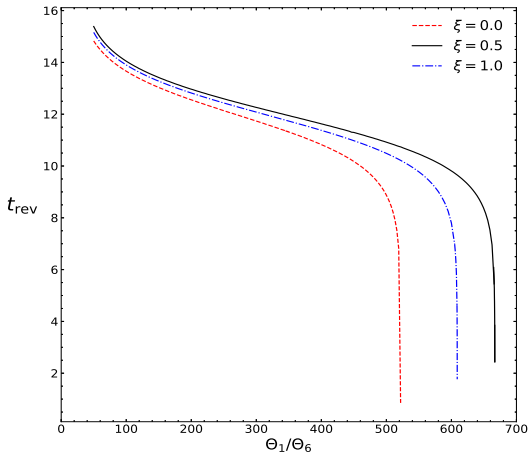}
\caption{Variation of $t_{\rm rev}$ as a function of $\Theta_1/\Theta_6$ for
expanding jets with $\vb=0.995,~p_1=p_6=0.1,~\rho_1=1.0$, but for composition $\xi=0$ (dashed, red), $\xi=0.5$ (solid,black)
and $\xi=1.0$ (dashed-dotted).}
\label{fig:revRS_xiexpJ}
\end{figure}
\subsection{Expanding jet}
\label{subsec:expndjet}
As the jet ploughs through the ambient medium, the jet head cross-section may expand. To employ the expansion effect we 
assume that the area across JH varies in time given by a function  
\begin{equation}
\label{eq:area_form}
A_4=1+C\frac{exp(t-t_0)}{1+exp(t-t_0)}
\end{equation}
Where $C$ is a constant which sets up and upper limit on $A_4$ and $t_0$ limits the onset time for expansion.
Figure (\ref{fig:profiles_jet}a) is plotted for $C=20.0~~\mbox{and}~t_0=15.0$.
From equation \ref{eq:pbal_var_area} one can conclude that across the JH
\begin{equation}
p_3A_3=p_4A_4
\label{eq:exp_area_pressure}
\end{equation}
Hence the pressure for an expanding jet also exhibits a jump across the CD unlike the Riemann problem where the area across the
CD was same. It may however be noted that some of the additional effects like back flow of jet material and generation of
transverse flow
components from the region between RS and CD has not been considered for simplicity. 
Figure (\ref{fig:profiles_jet}a) shows the variation of area across JH ($A_4$) with respect to time $t$.
One may refer to Fig. (\ref{fig:cartun}),
where we have shown expansion across the jet head, 
we have assumed the beam
area $A_1$ to be unity and $A_6=A_4$, also the area across the reverse shock surface is same on both sides ($A_3=A_1$).
In all other sections, we have assumed invariant jet cross-section.
In Fig. (\ref{fig:profiles_jet}b) we have plotted the propagation velocities $\vrs$ (dashed-dotted, blue), $\vj$ (solid, red)
and $\vfs$ (dashed, black) for a pressure matched electron-proton jet
propagating in a homogeneous electron-proton ambient medium. The variation of pressure and density in post-shock region of FS
(region 4) with respect to time also shown plotted in panel (c) and (d) respectively. Initial discontinuity is at $x_0=0.25$ and
the initial flow parameters are given as  
\begin{equation}
\rho_1=1.0,~\rho_6=500.0,~p_1=p_6=0.1,~v_1=0.995,~v_6=0.0
\end{equation} 
Interestingly, as the jet slows down, $\vrs$ may become negative or in other words, the RS
may start to move backward, as is shown in Fig. (\ref{fig:profiles_jet}b), where the RS starts to move backward at
$t\sim 10.4$. The corresponding solutions ($\rho,~p,~v^x,~\&~\Gamma$) at different epochs are presented in
Figs. (\ref{fig:expanding_jet}a-d). The overall jet structure advances forward (to right), however at $t>10.4$
Fig. (\ref{fig:profiles_jet}b) shows that
$\vrs<0$. Therefore Fig. (\ref{fig:expanding_jet}c) shows that the RS moves forward from $t=0~\rightarrow~ 10.4$,
but it moves backward $t=10.4 \rightarrow 10.6$ (inset shows, solid-red curve for $t=10.6$ and dashed-dotted-blue for $t=10.4$).
Although, $FS$ and $CD$ continue to move forward. For the initial stages of jet propagation $(t\leq 8.0)$, from
Fig \ref{fig:profiles_jet}a one can see that the area across the jet head is not changing hence the solution is self similar which
is evident from Fig. \ref{fig:expanding_jet}.
It may however be noted that, different values of $p$ and initial density
contrast $\rho_6/\rho_1$ would give different values of $\vj$, $\vfs$ and $\vrs$.
It is interesting to note that certain physical condition of the
environment of the jet may cause the RS to revert back!
In Fig. (\ref{fig:revRSexpJ}a), we plot the time $t_{\rm rev}$ at which the RS starts to move in the reverse direction
as a function of initial ratio of $\Theta_1/\Theta_6$, for 
two electron-proton jets with the same injection Lorentz factor $\gamma_1=10$ (i. e., $\vb=0.995$)
but different pressures and densities of the jet beam $p=0.1,~\&~\rho_1=1.0$ (solid, blue) and $p=0.01~\&~\rho_1=0.1$ (open
circle, red).
Both of the jets are initially pressure matched with the ambient medium. So for a given $\Theta_1/\Theta_6$,
initial ambient density $\rho_6$ scales similarly
for both the jets but have different values. For example, the jet with $p=0.1$ (solid, blue), $\Theta_1/\Theta_6=300$ implies
$\rho_6=300$ while for the jet $p=0.01$ (open circle, red), the same $\Theta$ ratio implies $\rho_6=30$.
In this figure, we have considered $\Theta_1/\Theta_6$ ratio from 50 to 600. 
The two curves exactly match for the entire range.
So even if the pressure and the density of the jets are not similar, the RS reverts back at the same time
for a given $\Theta_1/\Theta_6$ ratio. It may be noted that, there exists a limiting value of $\Theta_1/\Theta_6$
($\sim 600$, in this particular case)
where the $t_{\rm rev}$ decreases drastically.
For any value of $\Theta_1/\Theta_6$ greater than this limiting value,
the RS will move opposite to the direction of propagation from the start.
In Fig. (\ref{fig:revRSexpJ}b), we compare $\rho$ as a function of $x$ of the jet cases, for parameters
corresponding to the black star on the curve
in Fig. (\ref{fig:revRSexpJ}a).
Although the density distribution differs between the two jets being considered here,
but the location of RS, CD and FS coincide. Therefore
even if the $\rho_1$ and $p$ are different for the same ratio of the $\Theta$, the jet evolution
is exactly similar. It may however be noted the with different $p$ and $\rho_1$, the jet kinetic luminosity $L_j$
differs along the curve. In contrast, if the composition parameter is varied for the jet and the beam,
$t_{\rm rev}$ is different even if $\vb$ and $\Theta_1/\Theta_6$ are the same. In Fig. (\ref{fig:revRS_xiexpJ}),
we plot the time of reversal of the RS $t_{\rm rev}$ with the ratio $\Theta_1/\Theta_6$ for jets
with $\vb=0.995,~p_1=p_6=0.1,~\rho_1=1.0$, however for jets for three composition parameter $\xi=0$ (dashed, red),
$\xi=0.5$ (solid, black) and $\xi=1.0$ (dashed-dotted, blue). Unlike Fig. (\ref{fig:revRSexpJ}a), the curves for each
composition are
different. This shows that the evolution of various structures of a jet crucially depend on $\xi$.

\begin{figure*}
\hspace{0.0cm}
\includegraphics[width=13cm,height=9cm]{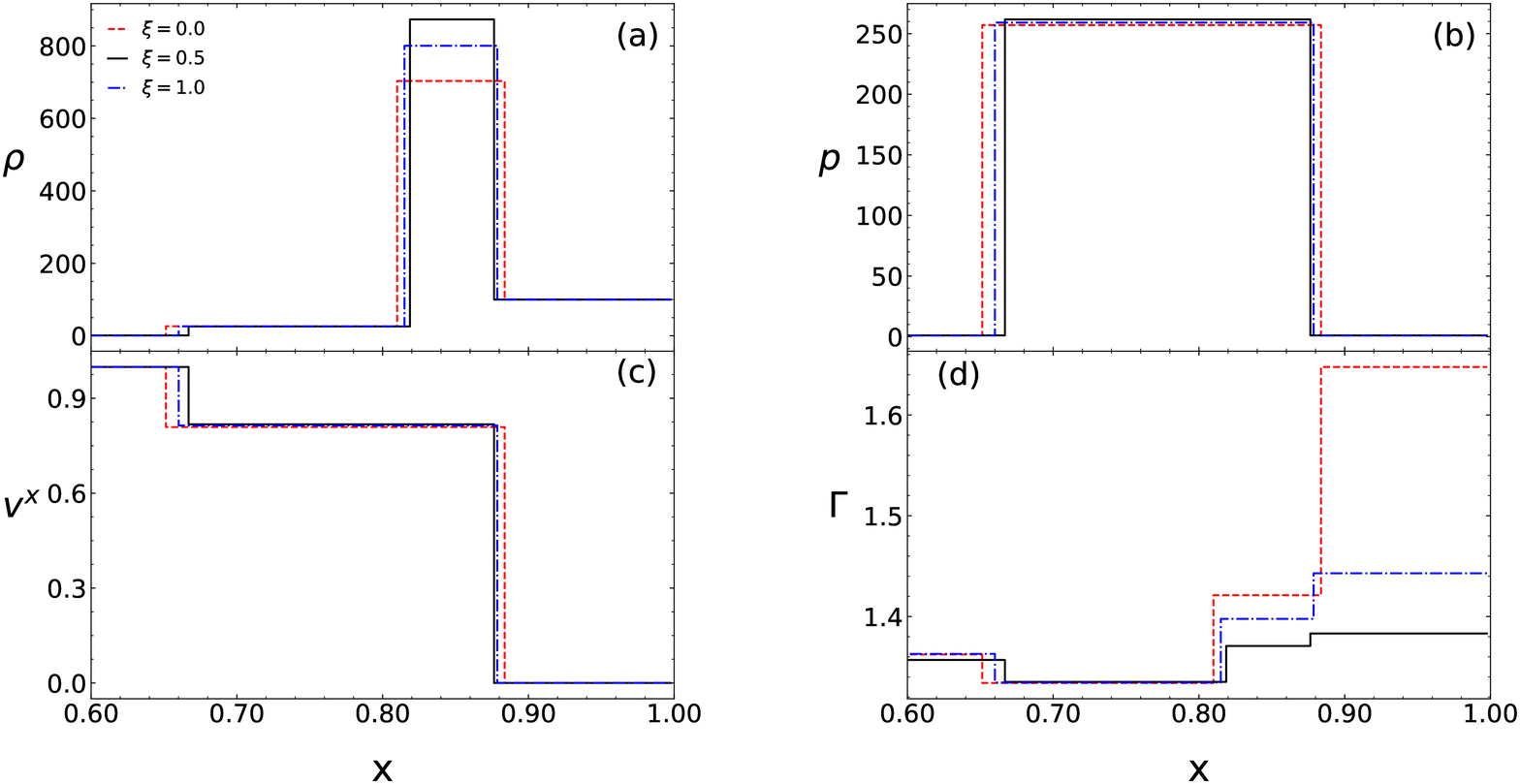}
\caption{Flow variables $\rho(a),~p(b),~v^x(c)\mbox{ and } \Gamma (d) $ as functions of $x$  at $t=1.0$.
The composition parameter of ambient medium is same as that of jet beam and mentioned in the legends in panel (a)	.
The initial condition is given by $\rho_1=1.0,~\rho_6=100.0,~p_1=p_6=1.0,~v_1=0.9988$ and $x_0=0.001$.}
\label{fig:1djsamexi}
\end{figure*}

\subsection{Effect of composition parameter}\label{subsec:compo}
The thermal state of the fluid depends upon the ratio $T /m$ ($m$ is the mass of the constituent particles
of gas) hence the change in composition
parameter will affect the thermal state.
To study the effect of the composition parameter on jet morphology we study two scenarios. In the first case, we keep the
composition parameter of ambient medium and jet beam same and for the second scenario, we study the case in which the
composition of ambient matter differs from the jet composition.
{To study the effect of composition we have assumed the uniform cross-sectional flow in all the cases}.

\subsubsection{Same composition parameter for jet and ambient medium}\label{subsec:samcompo}
Figure (\ref{fig:1djsamexi}) compares the solution snapshots of a relativistic jet at $t=1.0$ for $\xi=0.0$ (dashed, red),
$\xi=0.5$ (solid, black) and $\xi=1.0$ (dash-dotted, blue),
with initial parameters
\begin{equation}
\rho_1=1.0,~\rho_6=100.0,~p_1=p_6=1.0,~v_1=0.9988
\label{eq:jetsamexi}
\end{equation}
The density shell between the jet head/contact discontinuity and forward shock is tallest for the jet beam with $\xi=0.5$
also the adiabatic index is
lowest for the same. Hence the jet beam which consists of electron, positron, and proton plasma is hotter than the pure
electron-positron jet and the electron-proton jet, provided that the initial injection parameters are same for all.
For all the cases,
pressure of the shocked material in between CD and FS is considerably higher than the pressure of ambient medium which leads to
the formation of
over pressured cocoons and these cocoons are responsible for high degree of collimation of the jets \citep{bc89}.
The region 4 of the flow with $\xi=0.5$ is denser than the flow solutions of the other two composition.
Even the pressure in region 3 and 4, of the flow with composition parameter $\xi=0.5$ is greater than that of the flow of other
compositions.  
Hence we expect the flow with $\xi=0.5$ to be more stable and maintain its collimation over a long-range than the
electron-proton jet or electron-positron jet.    

\subsubsection{Different composition for jet and ambient medium}\label{subsec:diffcompo}
\begin{figure*}
\hspace{0.0cm}
\includegraphics[width=13cm,height=9cm]{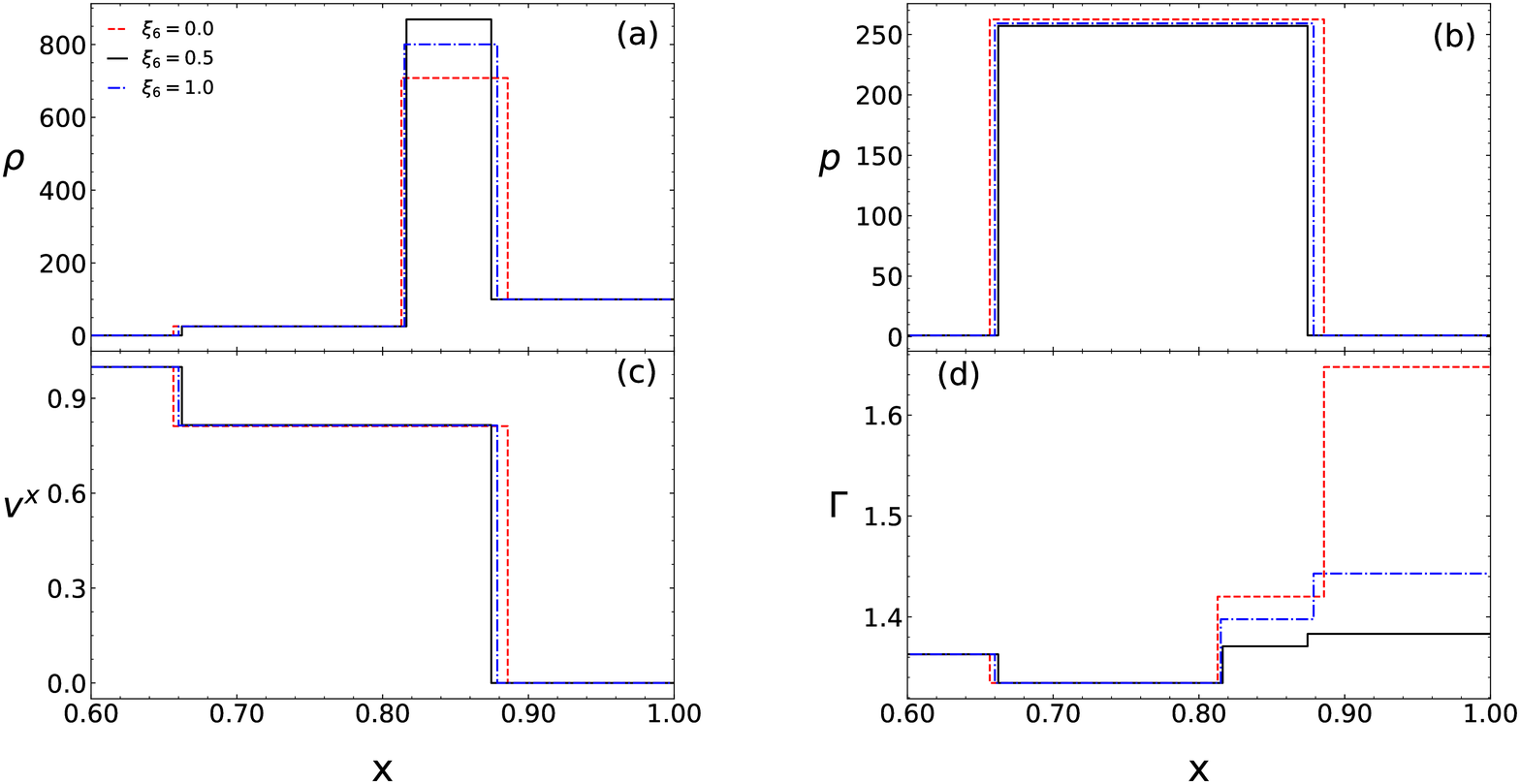}
\caption{Flow variables (a) $\rho$, (b) $p$, (c) $v^x$ and (a) $\Gamma$ as functions of $x$  at $t=1.0$ for an ambient medium
with composition parameter $\xi_6=0.0~\mbox{(red dashed)},~\xi_6=0.5~\mbox{(solid black)},~\xi_6
=1.0~\mbox{(blue dot-dashed)}$ and $x_0=0.001$.
The composition parameter of jet beam is taken to be $\xi_1=1.0$.}
\label{fig:1djdiffxi}
\end{figure*}

Figure (\ref{fig:1djdiffxi}) compares the snapshot of flow variables at $t=1.0$,
for the case when an electron-proton jet ($\xi_1=1.0$)
is injected into ambient medium with different composition parameters like $\xi_6=0$ (dashed, red), $\xi_6=0.5$
(solid, black) and $\xi_6=1.0$ (dash-dotted, blue). 
The initial flow parameters are
\begin{equation}
\rho_1=1.0,~\rho_6=100.0,~p_1=p_6=1.0,v_1=0.9988 \mbox{ and } \xi_1=1.0
\label{eq:diff_xi_param}
\end{equation}
The jet for which the ambient medium is $\xi_6=0$, has the widest and tallest high pressure region bounded by
FS and RS. It also has the widest high density shell (the region between CD and FS), although the height of the shell
is smallest compared to the other two cases.
The FS travels with the highest velocity for $\xi_6=0$ because of the low resistance offered by the medium.
The shock heating is maximum for electron, positron \& proton plasma resulting in the lowest adiabatic index in comparison to others. 
Since the jet beam initial conditions are exactly same in all the three cases, the ambient medium with $\xi_6=0$
is the least hot ($\Gamma \sim 5/3$) and therefore offers the least resistance to the jet, which causes the difference in the
jet structure as well.

\begin{figure*}
\hspace{0.0cm}
\includegraphics[width=13cm,height=9cm]{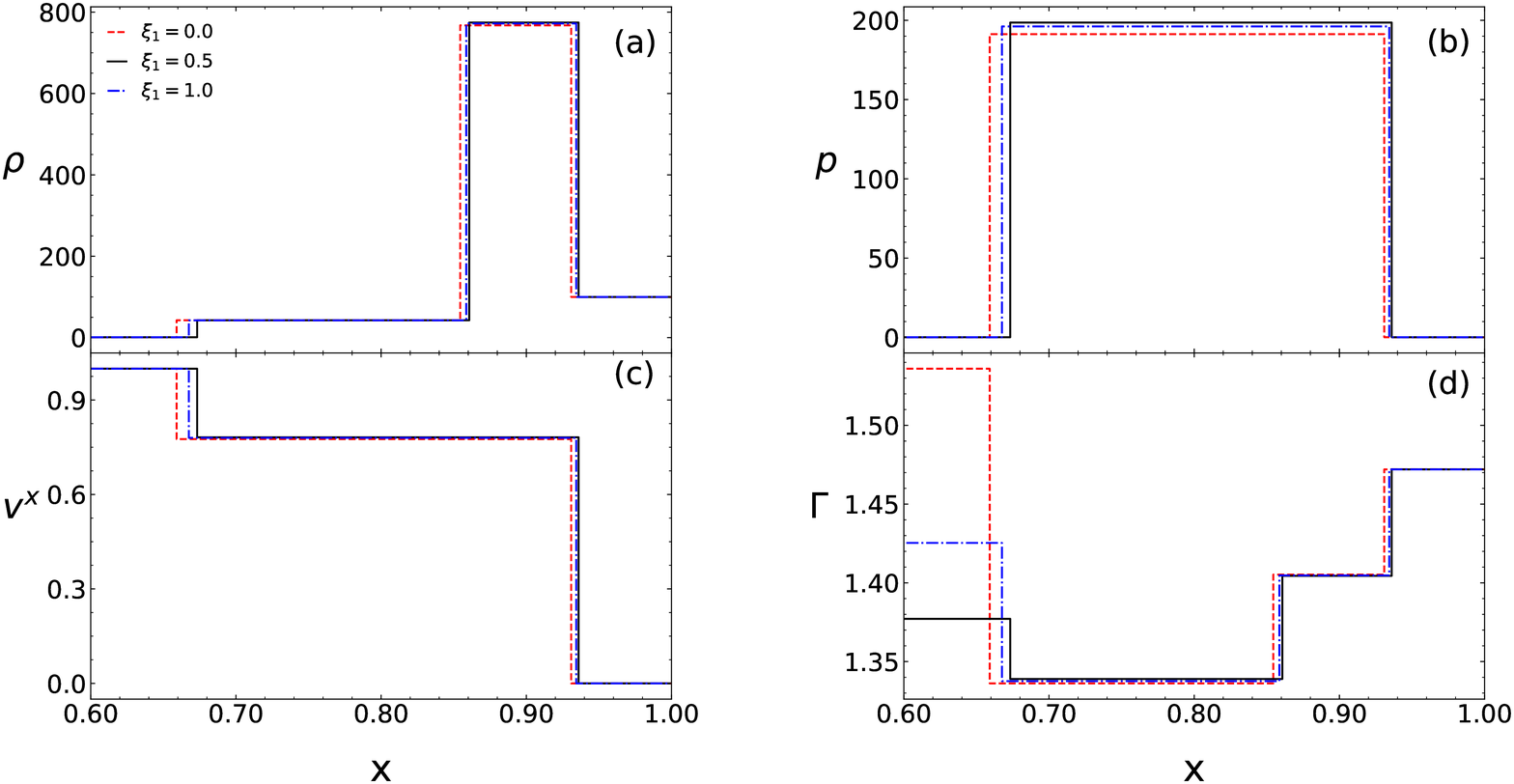}
\caption{Flow variables $\rho(a),~p(b),~v^x(c)\mbox{ and } \Gamma (d) $ as functions of $x$  at $t=1.0$ for a jet beam with
composition parameter $\xi_1=0.0~\mbox{(red dashed)},~\xi_1=0.5~\mbox{(solid black)},~\xi_1=1.0~\mbox{(blue dot-dashed)}$.
The composition parameter of ambient medium is taken to be $\xi_6=1.0$. The initial discontinuity is at $x_0=0.001$.}
\label{fig:diff_beam_xi}
\end{figure*}

We would now study the effect of composition of a jet beam flowing through the same ambient medium. We consider the cold
ambient medium of a static electron-proton fluid. Figure (\ref{fig:diff_beam_xi}) compares the flow variables for jets of different
composition traveling in the ambient medium composed of electrons and protons. The composition of the jet beam are
$\xi_1=0.0$ (dashed, red), $0.5$ (solid, black) and $1.0$ (dash-dotted). The injected Lorentz factor of the jet is 30
and the initial conditions are,
\begin{equation}
\rho_1=1.0,~p_1=0.1,~\rho_6=100.0,~p_6=0.1,~v_6=0.0 \mbox{ and } \xi_6=1.0
\end{equation}
Since the initial condition of the jet is the same, therefore the height of the high density shell between FS and CD have somewhat
similar values. Even the height of the high pressure region between RS and FS also is similar. Although
pair plasma jets has the lowest $\rho$ and $p$ values. 
The difference in RS and FS velocities determines the size of post shock region. The post shock region in blazar jets is considered
as the flaring region. This region is broadest for electron-positron jet.      

\begin{figure*}
\hspace{0.0cm}
\includegraphics[width=13cm,height=9cm]{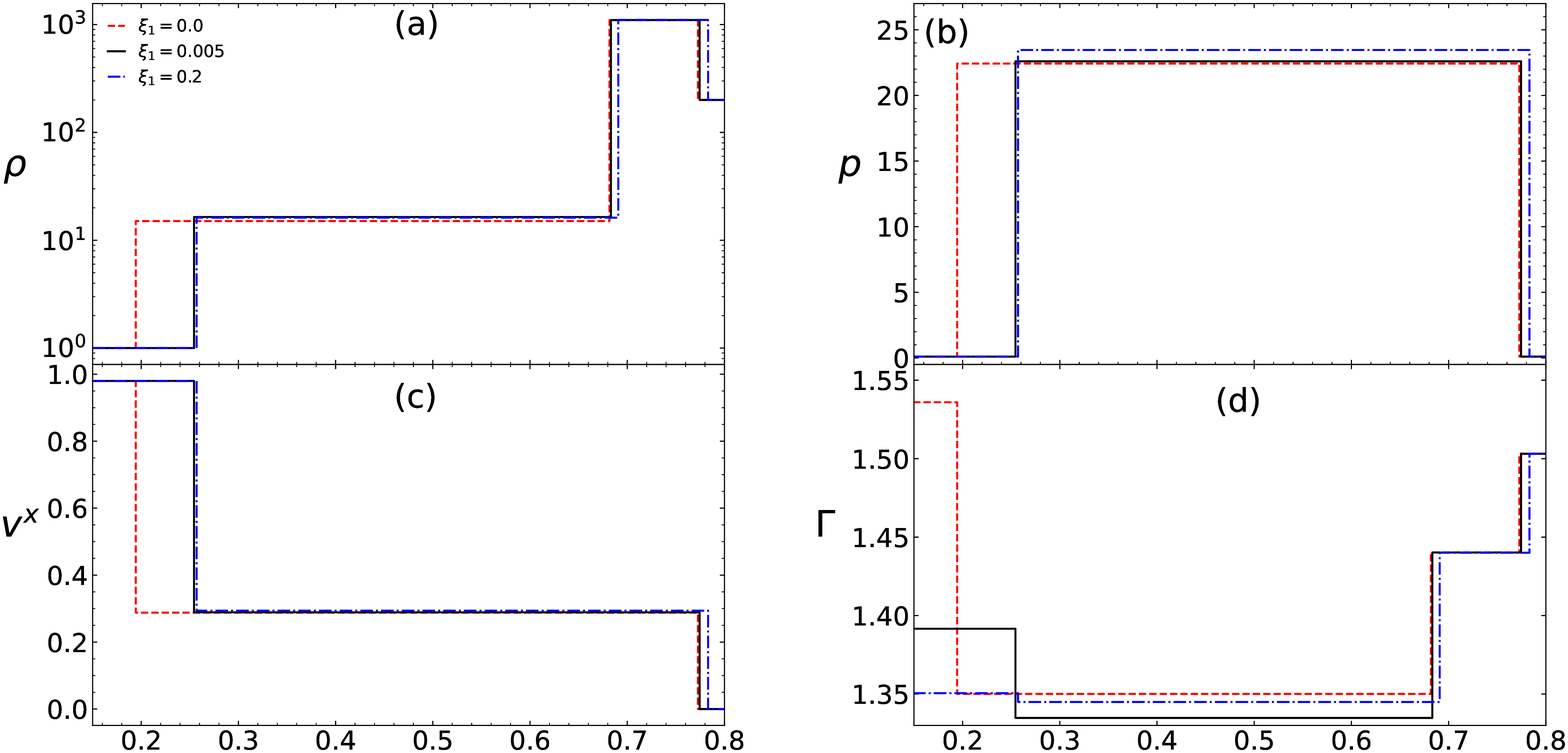}
\caption{Flow variables $\rho(a),~p(b),~v^x(c)\mbox{ and } \Gamma (d) $ as functions of $x$  at $t=1.0$ for a jet beam with
different composition parameters. The ambient medium consists of pure electron-proton plasma.}
\label{fig:rs_transition_xi}
\end{figure*}
\subsubsection{Effect of composition on reverse shock}\label{subsec:onrs}
\begin{figure*}
\hspace{0.0cm}
\includegraphics[width=13cm,height=5cm]{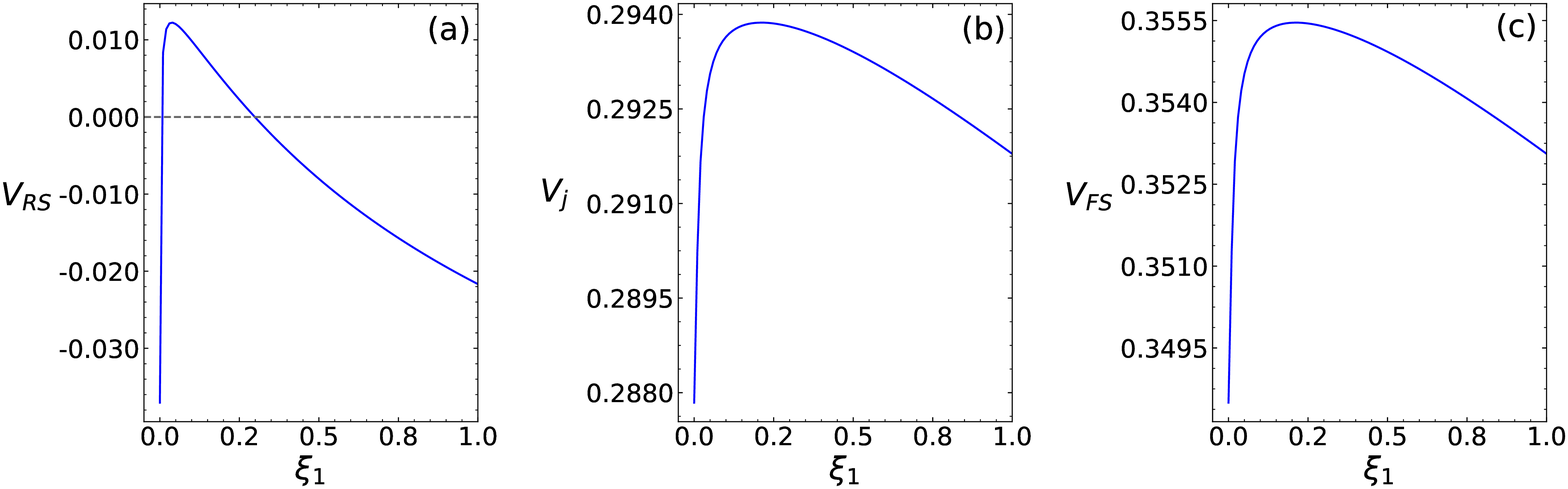}
\caption{The propagation velocities of (a) the RS i. e., $\vrs$, (b) the CD i. e., $\vj$ and (c) the FS i. e., $\vfs$
are plotted as a function of $\xi$ at $t=1.0$ for a jet beam with different composition parameters.
The ambient medium consists of pure electron-proton plasma. Initial conditions same as Fig. \ref{fig:rs_transition_xi}.}
\label{fig:vel_xi}
\end{figure*}
\begin{figure}
\hspace{0.0cm}
\includegraphics[width=\columnwidth]{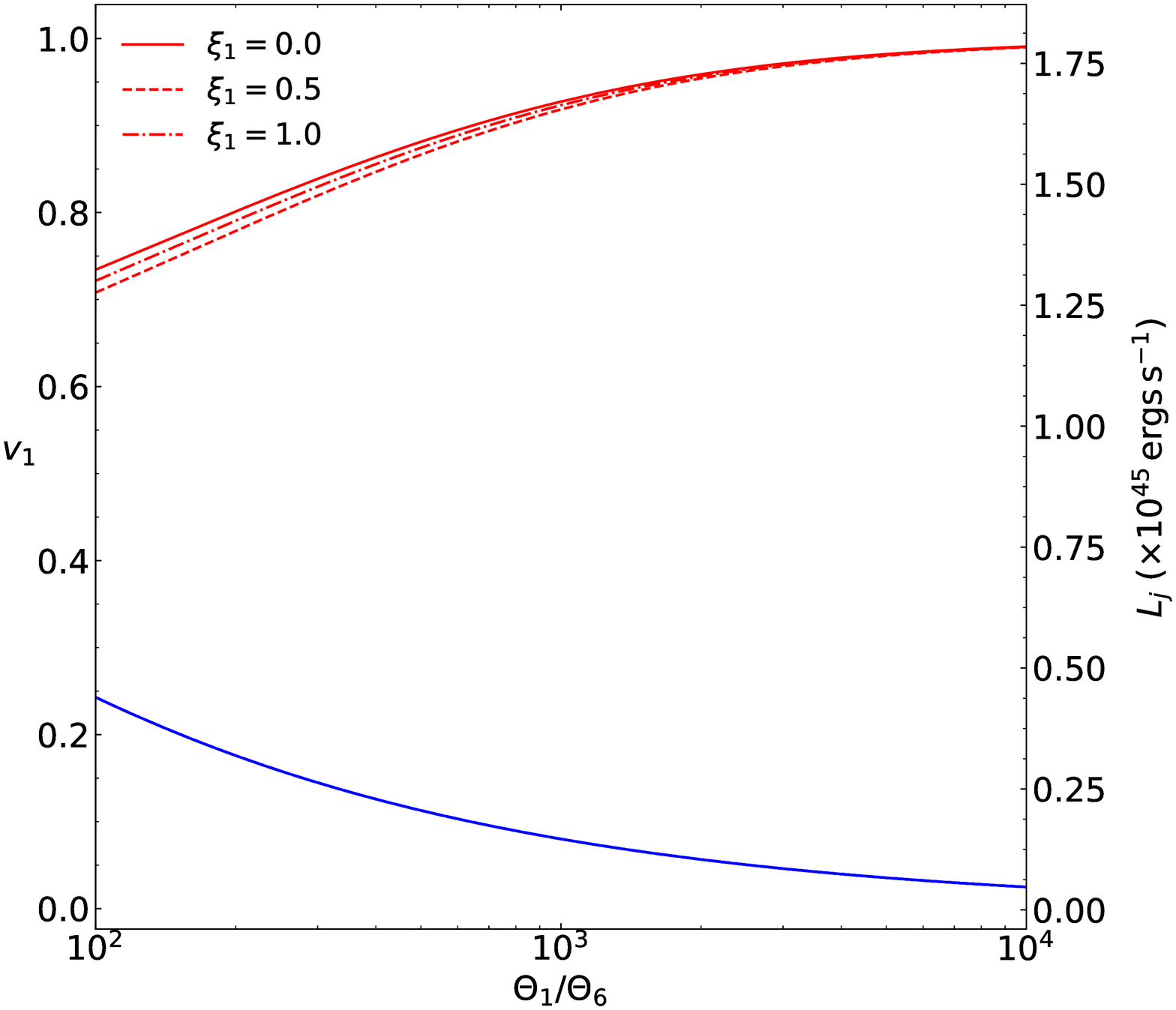}
\caption{Injection speed $\vb$ ($L_j$) as a function of $\Theta_1/\Theta_6$, for $\rho_1=1.0$,
$p_1=1.0$ and $p_6=0.1$. The curves corresponds beam composition $\xi_1=0$ (solid), $\xi_1=0.5$ (dashed) and
$\xi_1=1.0$ (dashed-dotted), while the composition of the ambient medium is $\xi_6=1.0$. The top curves (red)
are upper value of $\vb$ (or $L_j$) for which the RS goes in the opposite direction to the JH
and the bottom curves (blue and merged for all
$\xi_1$) represent the upper limit of $\vb$ (or $L_j$) for which RF is formed instead of RS.}
\label{fig:diffBmxiStruct}
\end{figure}
The composition of the jet beam can also affect the direction of propagation of the reverse shock.
In Fig. (\ref{fig:rs_transition_xi}) we compare the flow variables of a pressure matched jet beam with
Lorentz factor of 10 injected into an electron-proton ambient medium. The initial flow parameters are given as
\begin{equation} 
\rho_1=1.0,~p_1=0.1,~\rho_6=100.0,~p_6=0.1,~v_6=0.0
\end{equation} 
The initial discontinuity was at $x=0.25$. Figure (\ref{fig:rs_transition_xi}) shows that for pure electron-positron jet beam
i.e. $\xi_1=0.0$ the reverse shock propagates towards the jet base while a small fraction of protons in the jet beam can
revert the direction of reverse shock propagation. In order to drive home this point, we plot the
$\vrs$ (Fig. \ref{fig:vel_xi}a), $\vj$ (Fig. \ref{fig:vel_xi}b) and $\vfs$
(Fig. \ref{fig:vel_xi}c) with the composition parameter of the jet beam $\xi_1$. For these flow parameters, the reverse
shock
moves in the opposite direction for the pair plasma jet. However, $\vrs$ increases rapidly with the addition of protons and maximizes
for $\xi_1 \sim 0.04$. $\vrs<0$ again for $\xi_1 \gsim0.24$. The effect of composition on various propagation velocities is quite clear,
although the effect is more pronounced on $\vrs$. Figure (\ref{fig:vel_xi}b) shows that for these parameters
the electron-positron jet is not the fastest jet, instead, a jet
with composition $\xi_1 \sim 0.18$ is the fastest.

Finally, in Fig. (\ref{fig:diffBmxiStruct}), we plot the injection speed $\vb$ as a function of $\Theta_1/\Theta_6$
for jets of various jet composition $\xi_1=0$ (solid, red), $\xi_1=0.5$ (dashed, red) and $\xi_1=1.0$ (dashed-dotted, red).
The lower curve
(blue) signifies the locus of injection velocity $\vb$ for different $\Theta_1/\Theta_6$, below which rarefaction fan (RF) form,
and above which reverse shock RS forms but with negative velocity. The upper curve represents the locus of the injection
speed above which the RS moves with positive velocity. Assuming the width of the beam $y_1$ as 1 pc, and the $\rho_1$ as
$10^{-26}{\rm g}~{\rm cm}^{-3}$, we used equation (\ref{eq:jetlum}) to map the jet luminosity $L_j$ (right label) from the values of
$\vb$. The limiting velocity $v_{\rm lim}$ does not depend on $\xi_1$ which is the lower curve and therefore coincides for all
$\xi_1$. However, the upper curve, which is the limiting injection speed for $\vrs>0$ depends on the composition of the jet.

\section{Discussion and Conclusion}
\label{sec:conclude}

In this paper, we obtained the exact solution of time-dependent, one-dimensional relativistic jets plying through external
media (or ambient media), by solving equations for the ideal special relativistic fluid composed of dissimilar particles and
the thermodynamics of the fluid governed by a relativistic equation of state. The relativistic EoS we used is called the CR EoS,
is an approximation of the Chandrasekhar EoS. It is difficult to obtain a general Riemann solution with a general EoS and has
been discussed in the introduction. For the sake of completeness, we presented the solution methodology and the solutions
for three Riemann problems --- relativistic shock tube, shock-contact-shock problem and the wall-shock problem
in the appendix.
The intended impact of this work is firstly to provide exact solutions for numerical simulation
codes to compare their results with. But more than that, we studied various properties of the time-dependent relativistic jets,
which are in general glossed over. We addressed the issue of the properties of reverse shock. When a reverse-shock will
form and when will it not. When the reverse shock move in the opposite direction to the jet head and when along with it.
We wanted to study the effect of expanding jet-head cross-section on the jet solution. We also wanted to study the effect
of the composition of the jet or the ambient medium on the jet solutions and on all these counts we did get interesting results.
In the process, we modified the method to solve for jet-like Riemann problem to incorporate the cross-section effect across
the jet-head. However, it should be noted that many features of a relativistic jet, like generation of transverse velocity
near the jet head, back flow etc are not incorporated in this model for simplicity. Such omission may produce a higher
estimate of jet propagation speed, although a suitable
modeling of the expanding jet geometry may produce reasonable estimates. The change in cross-section across
the jet-head is an approximate attempt to mimic the tranverse expansion of the jet and has been discussed
in one part of this paper (section \ref{subsec:expndjet}). Such modifications across
CD have been attempted earlier \citep{bc89,myt04}.

In this paper, we obtained the condition for the formation of reverse shock and showed that it crucially depends on the injection
speed and the initial density contrast between the jet and the ambient medium, although for pressure matched jet that
critical injection speed turns out to be zero, in other words, in a pressure matched jet a reverse shock will always form.
However, if the jet-head cross-section is larger than the beam cross-section, then the reverse shock may travel in the
reverse direction. For the particular case of a pressure matched jet which is launched with a Lorentz factor $10$ with an
initial density contrast of $500$, if the cross-section across the jet head remains invariant for
some initial time and then expands for some time and settles into a constant value, then
we showed that initially RS, CD, and FS moved
in the positive direction, after a code time $t\gsim 10.4$, RS reversed back. And then as the cross-section settles
to a certain uniform value, then
the RS also settles to a uniform negative velocity. 
We also showed that the time of reversal of RS $t_{\rm rev}$ is unique for a given
ratio of the initial $\Theta$ of the beam and the ambient medium, if the injection speed
and composition is same. However, for different values of $\xi$, $t_{\rm rev}$ will be different even if $\vb$ and initial $\Theta$ 
ratio is same. Therefore, $\xi$ affects the jet evolution.
In general, the propagation speed depends on the beam injection velocities, jet-head cross-section as well as
initial density contrast across the surface of separation. 
We also showed that the composition of the jet also affects the jet structure as it evolves in time, even if all the initial
macroscopic fluid variables are same. 
The composition of the jet plasma affects the jet flow variables like density and pressure and also the propagation
velocities like $\vrs$, $\vj$ and $\vfs$ too. 
If jets of different composition flow through an ambient medium of a certain composition, then the jet with
less baryons are pushed back. Depending on initial conditions if the jet beam and jet head cross-sections are different then, we showed that for both baryon poor and baryon loaded jets,
the reverse shock can go back, even though the jet-head is moving in the positive direction. The composition of the jet not only
quantitatively affect the jet flow like different density, pressure and velocity distribution, but can also produce qualitatively
different effects like shocks moving in opposite direction for one composition and shocks
moving in the same direction for another composition. We also estimated the injection speed for an initial pressure
mismatch jet, for which rarefaction fan (RF) will form instead of the reverse-shock RS and also the upper limit of injection speed
for which the RS has negative velocity. Assuming the jet cross section of the order of 1 pc, these injection speeds correspond to a
jet kinetic luminosity of around $10^{45}$erg s$^{-1}$. Interestingly this upper limit of injection speed depends on the composition
parameter even for uniform cross-section jet. Our results indicate that in FR II jets typically with jet kinetic luminosities above
$10^{46}$ erg s$^{-1}$, both FS and RS would move forward. On the other hand,  in FR I jets with lower luminosities, either  backwardly
moving RS or RF could form. Hence, the numerical simulations of  FR I jets need special care, partly because of it,
but also because other processes including the entrainment of ambient material seems to become important \citep[e. g.,][]{pm07}.

\section*{Data Availability}
The data underlying this article will be shared on reasonable request to the corresponding author.

\section*{Acknowledgments}
The work of DR was supported by the National Research Foundation (NRF) of Korea through grants 2016R1A5A1013277 and
2020R1A2C2102800. RKJ acknowledges Mr.
Kuldeep Singh for help in python and cartoon diagram plotting.

\appendix
\section{Exact solution of Riemann Problem with CR EoS}\label{app:riem}
Riemann problem is the time evolution of an initial discontinuity between two non-uniform
states.
Depending upon the initial conditions it can evolve into different combinations of rarefaction and shock waves and the possible
scenarios are 
\begin{enumerate}
\item Rarefaction-Contact-Shock (RCS)
\item Shock-Contact-Shock (SCS)
\item Rarefaction-Contact-Rarefaction (RCR)
\end{enumerate} 
Exact solutions of Riemann problem have been extensively used to test
the accuracy of numerical simulation codes.
Here, we present solutions for three type of Riemann problems,
(i) RCS (as shown in Fig. \ref{fig:schem}b in the main text) which is known as
the shock-tube test, (ii) SCS which is mathematical model of a situation when two fluids collide head on supersonically,
and (iii) wall-shock (WS) problem which is physically equivalent to a supersonic flow which hits a wall. We also
compare the analytical WS solution with that of a relativistic one-dimensional relativistic TVD code obtained
by \citet[][]{rcc06,crj13}. 
The solution to the Riemann problem is obtained by solving the jump condition across shock, and solving an ordinary differential
equation arising out of self similarity condition of the rarefaction waves.

\subsection{RCS}
The RCS problem has a rarefaction fan, a contact discontinuity and a forward shock.

\subsubsection{Shock front evaluation}
\label{sec:shokfront}
The evolution of shock wave is governed by Rankine-Hugoniot jump conditions as described in section (\ref{sec:shock}). The post shock velocities are obatined as
\begin{equation}
v_b^{y,z}=h_a\gamma_av_a^{y,z}\left[\frac{1-(v_b^x)^2}{h_b^2+(h_a\gamma_av_a^t)^2} \right]^{1/2}
\label{eq:transvel_stube}
\end{equation}
Where subscript $a\,(b)$ denotes the state ahead (behind) the shock and $v^t$ is the absolute
value of tangential velocity of the flow. 
\begin{equation}
v^t=\sqrt{(v^y)^2+(v^z)^2}
\end{equation}
            
The normal component of ``star state'' velocity $v_s^{x*}$ prior to the shock front is by equation (\ref{eq:vbx}). 
\begin{equation}
v_s^{x *}=\frac{\left[h_a\gamma_av_a^x+{\gamma_s(p^*-p_a)}/{j} \right]}{\left[h_a\gamma_a+(p^*-p_a)
\left\{{\gamma_sv_a^x}/{j}+{1}/({\rho_a\gamma_a})\right\}\right]}
\label{eq:vxpostshk}
\end{equation}
Here $*$ and $a$ are the states behind and ahead to the shock front. For the right (left) shock, $a$ will be the initial right
(left) state.
As from equation (\ref{eq:masflux2})
\begin{equation}
j=j(\rho_a, p_a, \rho^*, p^*)
\end{equation}
we need the value of density in star state region $\rho^*$ in order to calculate $v_s^{x*}$. This value of density can be calculated
by solving the Taub adiabat (eq. \ref{eq:taubadiabat}) for $\rho^*$ using any iterative root finder method (we have used bisection method) for a given
value of $p^*$. Now we have to match this $p^*$ and $v_s^*$ with the starred pressure and velocity obtained from RF tail.

\subsubsection{Relativistic Rarefaction Waves}
Rarefaction waves are represented by the self-similar solutions of the flow equations. All the quantities describing the
fluid depend on the variable $\alpha=(x-x_0)/t$, where $x_0$ is the position of the interface separating the initial left and
right states \citep[see,][for details]{pmm00}. 
Substitution of the derivatives of $x$ and $t$ in terms of derivative of $\alpha$ (equation \ref{eq:eomconsv}) results in 
\begin{equation}
\begin{split}
& (v^x-\alpha)\frac{d\rho}{d\alpha}+\left\lbrace\rho\gamma^2v^x(v^x-\alpha)+\rho\right\rbrace\frac{dv^x}{d\alpha}\\
& +\rho\gamma^2(v^x-\alpha)\left(v^y\frac{dv^y}{d\alpha}+v^z\frac{dv^z}{d\alpha}\right)=0
\end{split}
\label{eq:dvxdbeta}
\end{equation}
\begin{equation}
\rho h\gamma^2(v^x-\alpha)\frac{dv^x}{d\alpha}+(1-v^x\alpha)\frac{dp}{d\alpha}=0
\label{eq:dpdbeta}
\end{equation} 
\begin{equation}
\rho h\gamma^2(v^x-\alpha)\frac{dv^y}{d\alpha}-v^y\alpha\frac{dp}{d\alpha}=0
\label{eq:dvydbet}
\end{equation} 
\begin{equation}
\rho h\gamma^2(v^x-\alpha)\frac{dv^z}{d\alpha}-v^z\alpha\frac{dp}{d\alpha}=0
\label{eq:dvzdbet}
\end{equation} 
From equation (\ref{eq:dvydbet}) and (\ref{eq:dvzdbet}) we conclude that if there is no tangential velocity in the initial state, no tangential flow will develop inside the rarefaction. Since the process along $\alpha$ is isentropic
\begin{equation}
\frac{dp}{d\alpha}=hc_s^2\frac{d\rho}{d\alpha}=\rho\frac{dh}{d\alpha}
\label{eq:adiab2}
\end{equation}
The determinant of the system (\ref{eq:dvxdbeta})-(\ref{eq:dvzdbet}) vanishes for the non-trivial solution, and one obtains either $\alpha=\beta_1$ or $\alpha= \beta_5$ (see, equation \ref{eq:eigenval}).
We are following the convention where the plus (minus) sign corresponds to the rarefaction wave propagating to right (left).
We can reduce the system (\ref{eq:dvxdbeta})-(\ref{eq:dvzdbet}) to an ordinary differential equation \citep{mm94,pmm00}
\begin{equation}
\rho h\gamma^2(v^x-\alpha)dv^x+(1-\alpha v^x)dp=0
\label{eq:ordeqraref}
\end{equation}
and two algebraic conditions
\begin{equation}
h\gamma v^y=\textrm{constant}
\label{eq:rarevy}
\end{equation}
\begin{equation}
h\gamma v^z=\textrm{constant}
\label{eq:rarevz}
\end{equation}
Equations (\ref{eq:rarevy}) and (\ref{eq:rarevz}) are similar to the equation (\ref{eq:transvel}) obtained for the shock.
Hence the expression for the transverse velocity given in equation (\ref{eq:transvel_stube}) can also be used to calculate the transverse
velocity components prior to the rarefaction waves.\\
    
Using the value of $\beta_{1,5}$ from equation (\ref{eq:eigenval}), equation (\ref{eq:ordeqraref}) can be written in the form
\begin{equation}
\frac{dv^x}{dp}=\pm\frac{1}{\rho h \gamma^2 c_s}\frac{1}{\sqrt{1+g(\alpha_\pm,v^x, v^t)}}
\label{eq:dvxdp}
\end{equation} 
\begin{equation}
\frac{dv^x}{d\rho}=\pm\frac{c_s}{\rho\gamma^2}\frac{1}{\sqrt{1+g(\alpha_\pm,v^x, v^t)}}
\label{eq:dvxdro}
\end{equation}

Where
\begin{equation}
g(\alpha_\pm,v^x, v^t)=\frac{(v^t)^2(\alpha_\pm^2-1)}{(1-\alpha_\pm v^x)^2}
\label{eq:gfactor}
\end{equation}

We have considered $\alpha_-=\beta_1$ and $\alpha_+=\beta_5$. The sign $\pm$ corresponds to the sign taken in equation (\ref{eq:eigenval}).
Normal velocity prior to the rarefaction can be calculated by integrating
the equation (\ref{eq:dvxdro}).
For ID or fixed $\Gamma$ EoS, the method to find out the solution of the Riemann problem is relatively easy. The equation of motion
describing the RF (equation \ref{eq:dvxdro}) admits an analytical solution for ID EoS, and is known as the Riemann Invariant. And therefore, starting from the left state (region 1), we use the Riemann Invariant to obtain the flow variables in region 3. While we use the shock conditions to obtain the flow variables
in region 4 in terms of those in region 6. Then equating velocity and pressure in region 3 and 4, we obtain a polynomial of the pressure.
Solving which we reconstruct the full solution. However, for a general EoS, equation (\ref{eq:dvxdro}) on integration does not admit an
analytical solution. Therefore, the solution of Riemann problem is not trivial for general EoS like CR.
In the following, we present the general method to solve the Riemann problem \citep[see,][for fluids with Newtonian equations
of motion]{k15}.

\subsubsection{Rarefaction Fan evaluation} \label{rarefaction}
For right going FS and left going RF (refer to Fig. \ref{fig:schem}b), it is clear that
the tail of RF (region 2) is adjacent to the starred state (region 3) and the head adjoins the initial state (region 1).
From Eq. (\ref{eq:adiab2})
we obtain the relation between $\rho, p,$ and $c_s$,
\begin{equation}
\frac{dp}{d\rho}=h(\rho, p)c_s^2
\label{eq:soundpeed2}
\end{equation}

To obtain the $p^*$ starting from initial state of region 1, equation (\ref{eq:soundpeed2}) is numerically integrated from initial
state with known pressure and density
to the final star state with given $p^*$ and unknown density $\rho^*$. The problem of determining
the unknown star state density is addressed
computationally by following the steps mentioned below,
\begin{enumerate}
 \item We solve the differential equation by RK-4 method for a constant density step size $\delta\rho$ for M number of integration
steps such that      
\begin{equation}
\bar{\rho}^*=\rho^M=\rho^{M-1}+\delta\rho
\label{eq:rostar1}
\end{equation}
where $\bar{\rho}^*$ is the provisional value of star state density. As the correct value of $\delta\rho$ is unknown, consequently we
do not know the correct value of $\rho_*$.
\item If $I(\delta\rho, \rho_a, p_a)$ be the value of pressure obtained by numerical integration then this value of pressure should be
equal to $p^*$, as the pressure remains  continuous across the contact. 
\begin{equation}
p^*-I(\delta\rho, \rho_a, p_a)=0
\label{eq:temp_p}
\end{equation} 
We can calculate the actual step size by solving the equation (\ref{eq:temp_p}). 
Once the correct value of step size is known the normal component of the flow velocity adjacent to the tail is calculated by integrating
the equation (\ref{eq:dvxdro})
\begin{equation}
(v^x _r)^*=v_a \pm \int_{\rho_a, p_a}^{\rho^*, p^*} \frac{c_s}{\rho\gamma^2}\frac{1}{\sqrt{1+g(\alpha_\pm,v^x, v^t)}} d\rho
\label{eq:vxrstar}
\end{equation}
And the corresponding spatial location for the $m^{th}$ integration step is given by
\begin{equation}
x^{(m)}=x_0+\alpha_{\pm}(\rho^{(m)}, p^{(m)}, v^{(m)})t
\label{eq:xm}
\end{equation}\\
The final step of integration ($M^{th}$ step) for equation (\ref{eq:xm}) corresponds to the position of tail.
Here $\alpha_+=\beta_5$ and $\alpha_-=\beta_1$
(from equation \ref{eq:eigenval}).
\end{enumerate}

As the normal component of the flow velocity across the contact discontinuity is continuous
\begin{equation}
(v^{x*} _r)-(v^{x*} _s)=0
\label{eq:velbalnc}
\end{equation}

The value of ``star state'' pressure is obtained by solving equation (\ref{eq:velbalnc}) by any iterative root finder method.
Once $p^*$ is computed the 
tangential velocity components can be calculated using equation (\ref{eq:transvel2}) for shock wave and
equations (\ref{eq:rarevy}, \ref{eq:rarevz}) for the rarefaction wave.

\begin{table*}
  \begin{center}
  \begin{tabular}{|l|l|l|l|l|l|l|l|l|l|l|l|l|l|l|}
  \hline
     $v^t_1$ & $v^t_6$  & $p_*$ & $v^x_*$ & $\rho_3$ & $\rho_4$  & $v^t_3$ & $v^t_4$ &$V_s$ & $x_H$ & $x_t$ \\  
     \hline
     0.000 & 0.000 & 1.49 & 0.66 & 1.22 & 4.80 & 0.0000 & 0.0000 & 0.78 & 0.36 & 0.54\\ 
     \hline
     0.000 & 0.900 & 2.91 & 0.47 & 2.00 & 6.34 & 0.0000 & 0.6422 & 0.60 & 0.36 & 0.47 \\ 
     \hline
     0.000 & 0.990 & 5.64 & 0.23 & 3.27 & 8.32 & 0.0000 & 0.9141 & 0.33 & 0.36 & 0.41 \\  
     \hline
     0.900 & 0.000 & 0.41 & 0.34 & 0.48 & 2.53 & 0.9140 & 0.0000 & 0.55 & 0.43 & 0.55 \\ 
     \hline
     0.900 & 0.900 & 0.98 & 0.28 & 0.89 & 3.98 & 0.9242 & 0.7997 & 0.40 & 0.43 & 0.53 \\ 
     \hline
     0.900 & 0.990 & 2.91 & 0.17 & 2.00 & 6.34 & 0.9267 & 0.9501 & 0.24 & 0.43 & 0.49 \\  
     \hline
     0.990 & 0.000 & 0.17 & 0.13 & 0.25 & 1.45 & 0.9903 & 0.0000 & 0.39 & 0.48 & 0.52 \\  
     \hline
     0.990 & 0.900 & 0.29 & 0.11 & 0.38 & 2.08 & 0.9912 & 0.8753 & 0.24 & 0.48 & 0.51 \\  
     \hline
     0.990 & 0.990 & 0.92 & 0.09 & 0.86 & 3.87 & 0.9927 & 0.9779 & 0.14 & 0.48 & 0.50 \\  
     \hline
  \end{tabular}
\caption{Solution of Shock tube problem problem at $t=0.25$ in spatial domain $x\in [0,1]$  with initial data
$p_1=10.0$, $p_6=0.1$, $\rho_1=5.0$, $\rho_6=1.0$, $v^x_1=0.0$, $v^x_6=0.0$ . The location of the initial discontinuity is at $x_0=0.5$, $x_H$ and $x_T$ are the positions of rarefaction head and tail
respectively. In all cases we have taken composition parameter $\xi=1.0$.}
\end{center}
\label{tab:tab1}
\end{table*}

\begin{figure*}
\hspace{0.0cm}
\includegraphics[width=12cm,height=8cm]{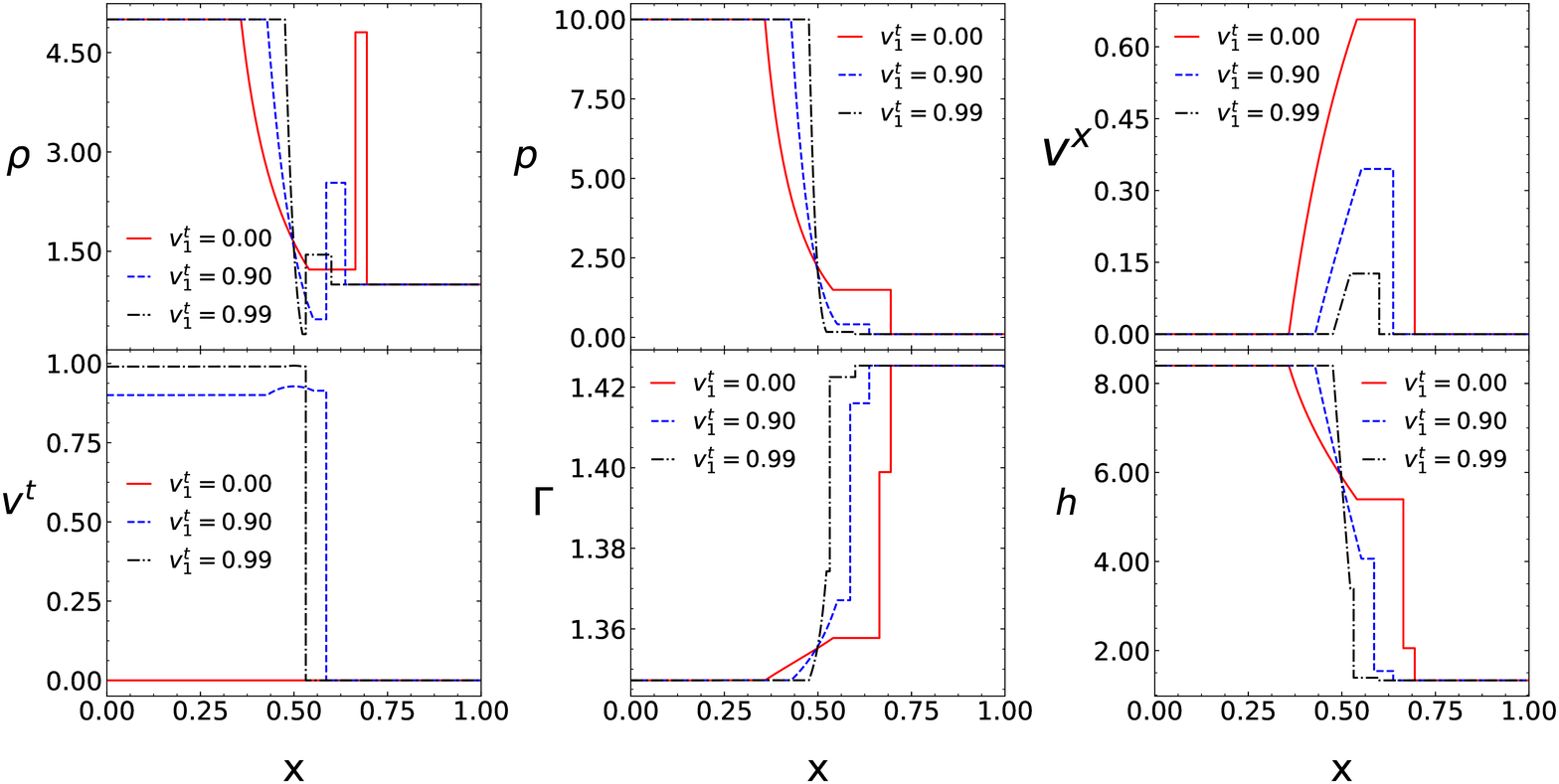}
  \caption{Comparison of $\rho$, $p$, $v^x$, $v^t$, $\Gamma$, and $h$, all plotted at the time $t=0.25$. Initial conditions $\rho_1=5,~\rho_6=1$, $p_1=10,~p_6=0.1$, $v^x_1=v^x_6=0.0$, $v^t_1=0$ (solid, red), $v^t_1=0.9$ (dashed, blue)
  and $v^t_1=0.99$ (dash-dotted, black) and for all cases initial
$v^t_6=0$. The fluid composition is $\xi=1.0$.}
\label{fig:stube_vt}
\end{figure*}

The solution of Riemann problem is given in table 1 and some of these solutions are shown in fig (\ref{fig:stube_vt}).
The existence of the upper limit of speed in relativity, causes various velocity components to be related to each other through the
Lorentz factor, 
solutions depend strongly on the
different combinations of initial $v^t$. For high values of $v^t$, the values of $v^x$
are low. 
Therefore, solutions with $v^t=0$ is faster and hotter than solutions with $v^t\neq 0$.
The solutions from relativistic TVD codes proposed before \citep{rcc06,crj13} have been compared with these Riemann solutions
and they agree well.

\subsection{Shock-Contact-Shock (SCS)}
\begin{figure*}
\includegraphics[width=12cm,height=8cm]{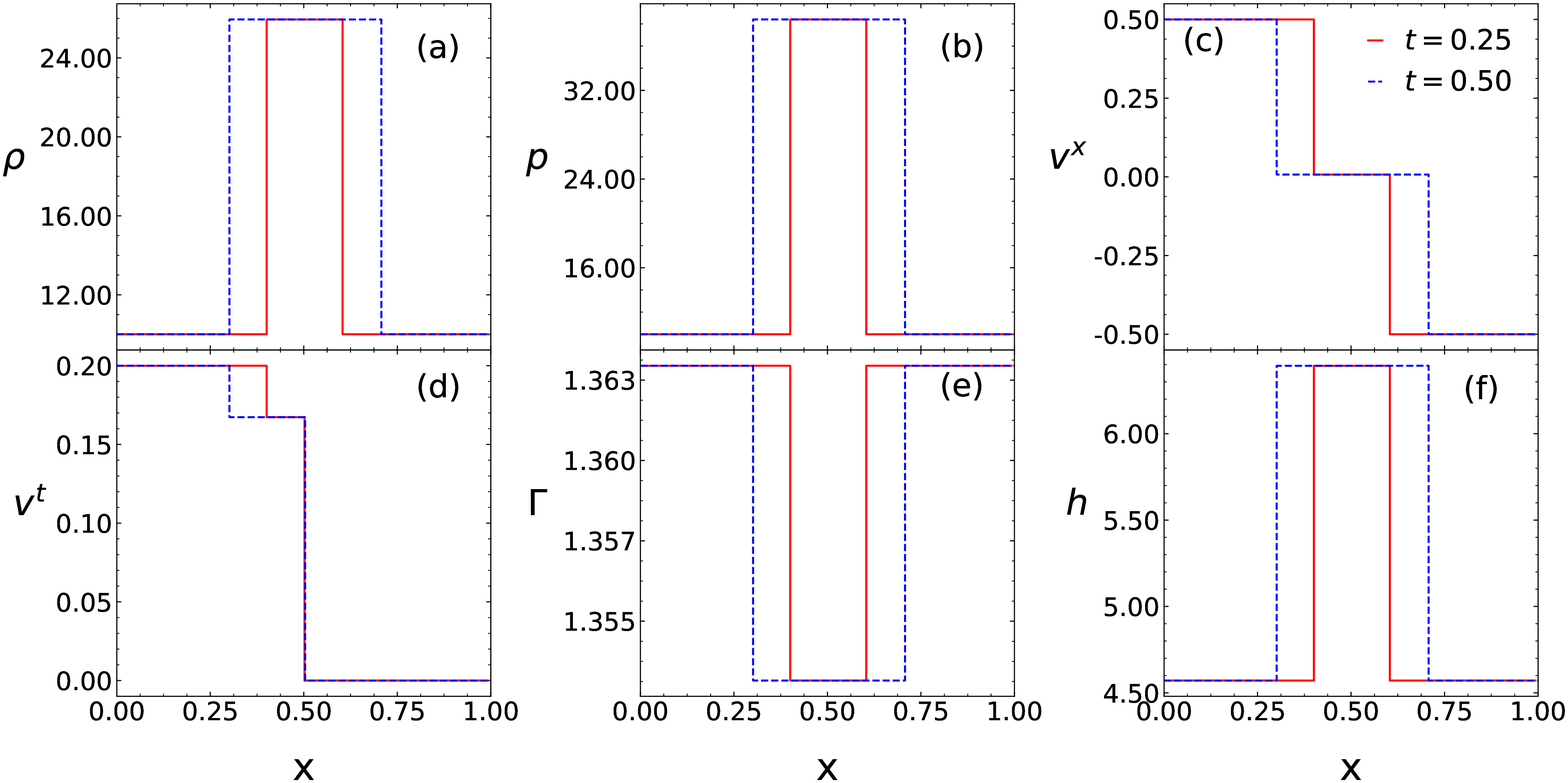} 
  \caption{SCS:Flow variables (a) $\rho$, (b) $p$, (c) $v^x$, (d) $v^t$, (e) $\Gamma$ and (f) $h$ as a function of $x$,
 for Shock-Contact-Shock case at $t=0.25$ and for fluid composition $\xi=1.0$.}
  \label{fig:Shock-Contact-Shock}
\end{figure*}

The collision of two streams is the physical scenario which can be associated with the shock-contact-shock (SCS) Riemann problem.
Referring back to of Fig. \ref{fig:schem}a, SCS can be represented if regions 2 and 5 both are surfaces of
the shock waves, instead of 2 being RF. Equation (\ref{eq:vxpostshk}) provides an expression for $v_3^x$ in terms of
$v_1^x$ for a left moving shock and an expression
for $v_4^x$ in terms of $v_6^x$ for a right moving shock.

\begin{equation}
\begin{split}
&v_3^{x}=\left(h_1\gamma_1v_1^x+\frac{\gamma_s(p_*-p_1)}{j} \right)\\
&~~~~~~~\left( h_1\gamma_1+(p_*-p_1)
\left( \frac{\gamma_sv_1^x}{j}+\frac{1}{\rho_1\gamma_1}\right)\right)^{-1}
\end{split}
\label{eq:v3scs}
\end{equation}

\begin{equation}
\begin{split}
&v_4^{x}=\left(h_6\gamma_6v_6^x+\frac{\gamma_s(p_*-p_6)}{j} \right)\\
&~~~~~~~\left( h_6\gamma_6+(p_*-p_6)\left( \frac{\gamma_sv_6^x}{j}+\frac{1}{\rho_6\gamma_6}\right)\right)^{-1}
\end{split}
\label{eq:v4scs}
\end{equation}
For equation (\ref{eq:v3scs}) $j$ is the negative root of equation (\ref{eq:masflux2}) and for equation (\ref{eq:v4scs}) it is
the positive root of equation (\ref{eq:masflux2}).
Across the contact discontinuity 
\begin{equation}
v_3^{x}-v_4^{x}=0
\label{eq:velbalnc_scs}
\end{equation}
Equation (\ref{eq:velbalnc_scs}) is solved for $p^*$ using the iterative root finder and rest of the quantities can be calculated
once $p^*$ is obtained. Densities in region 3 and 4 are obtained from the Taub's adiabat for left and right shock respectively.
The solution ($\rho,~p,~ v^x,~v^t,~\Gamma,~\&~h$) is presented in Fig. (\ref{fig:Shock-Contact-Shock}a-f) for two time snap
$t=0.25$ (solid, dashed) and $t=0.5$ (dashed, blue). It is clear that both the shocks are moving apart from each other
and the shock velocity is actually in the direction opposite to the local flow velocities $v^x$.
The initial condition used for this particular model is

\begin{equation}
\begin{split}
 \rho_1=10.0,~p_1=10,~v^x_1=0.5,~v^t_1=0.2;\mbox{ and } \\ \rho_6=10.0,~p_6=10.0,~v^x_6=-0.5,~v^t_6=0.
\end{split}
\label{eq:scs_ini}
\end{equation}
The two shock surfaces are at the either side of $x=x_0$, and the CD is marked by the jump in $v^t$.

\subsection{Solution of relativistic wall shock problem}
\label{sec:walsok}

The relativistic wall shock problem is a mathematical model of the collision between a fluid of density $\rho_1$
moving with extremely high velocity $v_1$ (presently along right) and a reflecting wall. 
The fluid after being reflected back, gets
compressed and eventually a reverse shock is generated. 
The shock moves in the opposite direction of the fluid leaving behind a hot
and compressed fluid with zero velocity in post shock region. The fluid density, pressure and velocity of fluid in post shock region are represented by $\rho_2$, $p_2$ and $v^x_2$ respectively. As velocity in post shock region $v^x_2=0$, we need only two equations to find out $\rho_2$ and $p_2$. We can use equations (\ref{eq:vbx}) and (\ref{eq:taubadiabat}) to find out the post shock density and pressure.
From equation (\ref{eq:vbx}) we have
\begin{equation}
h_1\gamma_1v^x_1+\frac{\gamma_s(p_2-p_1)}{j}=0 
\label{eq:wallshock_1}
\end{equation}
As shock is moving towards the left direction $j$ should be taken as negative root of equation (\ref{eq:masflux2})  
\begin{equation}
j=-\left(-\frac{(p_1-p_2)}{\left({h_1}/{\rho_1}- {h_2}/{\rho_2}\right)}\right)^{1/2}
\end{equation}
From equation (\ref{eq:taubadiabat})
\begin{equation}
h_1^2-h_2^2=\left(\frac{h_1}{\rho_1}+\frac{h_2}{\rho_2}\right)(p_1-p_2)
\label{eq:wallshock_2}
\end{equation} 
We can solve the equations (\ref{eq:wallshock_1}) and (\ref{eq:wallshock_2}) using the Newton-Raphson method to obtain
$\rho_2$ and $p_2$. The shock velocity is calculated using (\ref{eq:shockvel})
\begin{equation}
V_s=\frac{\rho_1^2\gamma_1^2v_1-|j|\sqrt{j^2+\rho_1^2\gamma_1^2(1-v_1^2)}}{\rho_1^2\gamma_1^2+j^2}
\label{eq:vs_sokwal}
\end{equation}
If $x_0$ is the location of wall then the shock location after time $t$ is given by
\begin{equation}
x_s=x_0+V_st
\label{eq:wallshockloc}
\end{equation}     

\begin{figure}
\includegraphics[width=\columnwidth]{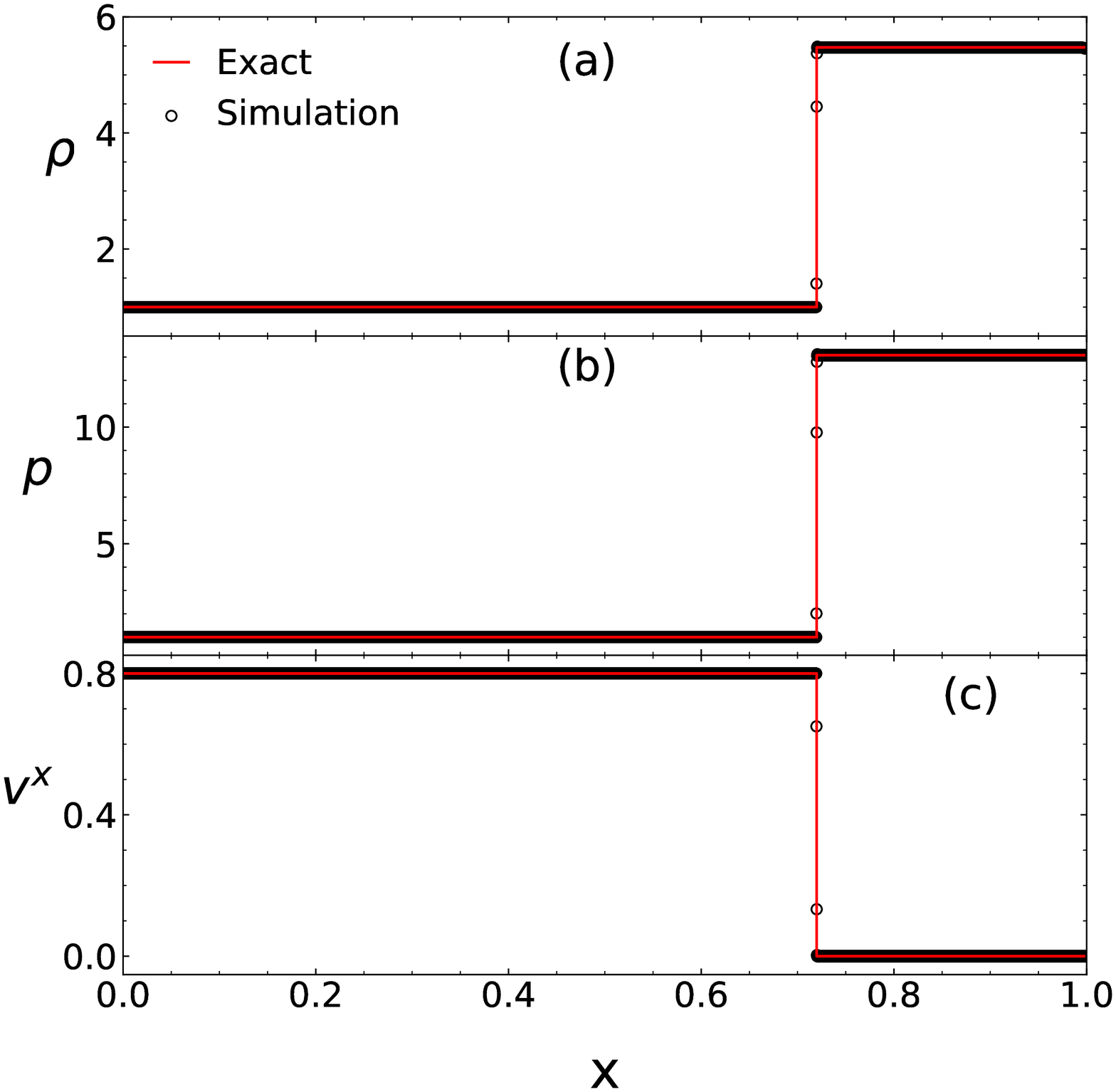} 
  \caption{WS: Solution of the wall shock problem with initial condition
  $\rho_1=1.0$, $p_1=1.0$, and $v_1=0.8$.
  The solution is obtained for wall located at $x_0=1.0$, composition parameter $\xi=0.5$ and $t=0.8$.}
\label{fig:wallshocksoln}
\end{figure}

In Fig. \ref{fig:wallshocksoln}a-c, we plot the flow variables $\rho$, $p$, and $v^x$ as a function of $x$. The solid red lines show the exact solutions and the solutions marked by black open circles are obtained using relativistic TVD code of \cite{crj13}.
The initial conditions for this problem are 
\begin{equation}
 \rho_1=1.0,~ p_1=1.0, \mbox{ and } v^x_1=0.8
\end{equation}

The solution contains only one discontinuity in the form of a shock jump, across which $\rho$, $p$ and $v^x$ are discontinuous.

\bsp	
\label{lastpage}
\end{document}